\documentclass[12pt,a4paper,oneside]{article}

\usepackage{graphicx,amssymb,amsmath}
\usepackage{setspace} % should be before hyperref, otherwise footnote links lead to first page
\usepackage[linkcolor={blue},citecolor={blue},colorlinks=true,linktocpage,urlcolor={blue}]
{hyperref}
\usepackage{cite}
\usepackage[margin=2cm]{geometry}
\usepackage[dvipsnames]{xcolor}
\definecolor{darkblue}{rgb}{0,0,0.54}
\usepackage{tikz}
\usetikzlibrary{patterns}
\usetikzlibrary{calc}
\usepackage{multirow} % Multi-column and multi-row cells \in table
\usepackage{appendix}
\usepackage{bbm}
\usepackage{import}
\usepackage{tensor}
\usepackage{pdfpages}
\usepackage{cancel} % when want to show that terms cancel, use \bcancel
\usepackage{slashed} % for Dirac slash notation
\usepackage{xparse} % for conditional newcommands (DeclareDocumentCommand)
\usepackage{tabu} % table
\usepackage{adjustbox} % fit to size, use in table
\usepackage{simpler-wick}
\usepackage[bb=boondox]{mathalfa}
\usepackage{dsfont}
\usepackage{subcaption}

\numberwithin{equation}{section} % number equations according to sections
\usepackage{empheq} % emphasizing equations

\graphicspath{{Figures/}} % in order to import figures from a subdirectory

\newcommand{\cO}{\mathcal{O}}

\newcommand{\cJ}{\mathcal{J}}

\newcommand{\pder}[2]{\frac{\partial#1}{\partial#2}} % leave blank if want only d/dy
\DeclareMathOperator{\sign}{sign}
\DeclareMathOperator{\tr}{tr}
\DeclareMathOperator{\li}{Li}

\newcommand{\JMS}{\mathbb{J}}

\usepackage{titling} % for a tit\le page
\usepackage[affil-it]{authblk} % for the institute etc in the title page

\usepackage{tocloft} % change table of contents font
\title{Semiclassical geometry in double-scaled SYK}
\author[1]{Akash Goel,}
\author[2]{Vladimir Narovlansky,}
\author[2]{Herman Verlinde}

\affil[1]{ Center for Cosmology and Particle Physics, New York University, New York, NY 10003, USA}
\affil[2]{ Department of Physics, Princeton University, Princeton, NJ 08544, USA}
\date{\small \texttt{akash.goel@nyu.edu, narovlansky@princeton.edu, verlinde@princeton.edu}}

\makeatletter % use numbers instead of symbols in the thanks part in the title
\let\@fnsymbol\@arabic
\makeatother

\begin{document}

\begin{titlingpage}
    \maketitle
    \begin{abstract}
	We argue that at finite energies, double-scaled SYK has a semiclassical approximation controlled by a coupling $\lambda $ in which all observables are governed by a non-trivial saddle point. The Liouville description of double-scaled SYK suggests that the correlation functions define a geometry in a two-dimensional bulk, with the 2-point function describing the metric. For small coupling, the fluctuations are highly suppressed, and the bulk describes a rigid (A)dS spacetime. As the coupling increases, the fluctuations become stronger. We study the correction to the curvature of the bulk geometry induced by these fluctuations. We find that as we go deeper into the bulk the curvature increases and that the theory eventually becomes strongly coupled.
    In general, the curvature is related to energy fluctuations in light operators.
    We also compute the entanglement entropy of partially entangled thermal states in the semiclassical limit.
    \end{abstract}
\end{titlingpage}

\tableofcontents

\section{Introduction}

The double-scaled Sachdev-Ye-Kitaev (SYK) model is a quantum mechanical theory of $N$ Majorana fermions $\psi_i$, where $i=1, \cdots ,N$, satisfying $\{\psi_i,\psi_j\}=2\delta _{ij} $. The fermions interact all-to-all, in groups of $p$ fermions, and the couplings are taken to be random. In more detail, the Hamiltonian is given by
\begin{equation}
H = i^{p/2} \sum _{1 \le i_1<i_2<\cdots <i_p \le N} J_{i_1 \cdots i_p} \psi_{i_1} \cdots \psi_{i_p} .
\end{equation}
The theory is a large $N$ theory, where we take $N,p$ to be large, keeping fixed the quantity \cite{Erdos:2014zgc,Cotler:2016fpe,Berkooz:2018qkz,Berkooz:2018jqr}
\begin{equation}
\lambda  \equiv \frac{2p^2}{N} .
\end{equation}
It is common to denote in double-scaled SYK \cite{Berkooz:2018jqr}
\begin{equation}
q \equiv e^{-\lambda} .
\end{equation}

For simplicity, we will take the couplings $J_{i_1 \cdots i_p} $ to be independent Gaussian variables with zero mean values. In this paper we will normalize the variance such that
\begin{equation}
\label{gaussianj}
\langle J_{i_1 \cdots i_p} ^2\rangle _J = \frac{1}{\lambda } \binom{N}{p} ^{-1} \JMS^2 \approx \frac{(p-1)!}{2pN^{p-1} } \JMS^2
\end{equation}
where the angular brackets denote the average over the couplings, and there is no summation over the indices $i_1, \cdots, i_p$ in the equation above. This normalization of the variance is the one compatible with \cite{Maldacena:2016hyu}.\footnote{In this reference, $\JMS$ is denoted by $\cJ$, while we reserve $\cJ$ for a different quantity.} It is related to the more conventional $\cJ$ used in double-scaled SYK, that is particularly convenient in the chords construction, via
\begin{equation}
\JMS^2= \lambda \cJ^2.
\end{equation}

Correlation functions have been calculated exactly in double-scaled SYK \cite{Berkooz:2018qkz,Berkooz:2018jqr} for any finite $\lambda $ (or equivalently, $q$). 
One expects the studies of the SYK model in which $p$ is not double-scaled with $N$ as above, but is rather independent of $N$ \cite{Sachdev:1992fk,Kitaev_talk,Polchinski:2016xgd,Maldacena:2016hyu}, to be related to the $\lambda  \to 0$ limit of double-scaled SYK. In fact, it was shown in \cite{Berkooz:2018jqr} that for $\lambda \to 0$ and restricting to low energies, the correlation functions found in double-scaled SYK agree precisely with the (quantum) correlators obtained in the Schwarzian theory \cite{Mertens:2017mtv} (for a recent review, see \cite{Mertens:2022irh}).

In this paper we study the small $\lambda $ limit, without restricting to low energies. One expects this limit to be related to the so called `large $p$' limit of SYK, where $p$ is taken to be large, but independent of $N$ \cite{Maldacena:2016hyu} (see also \cite{Streicher:2019wek,Choi:2019bmd}). Indeed, at leading order we reproduce exactly the results of large $p$ SYK. We observe that the exact results for the various observables of double-scaled SYK become classical in the $\lambda  \to 0$ limit. That is, the integrands become sharply peaked and dominated by a saddle point. The large $p$ results are simply the classical value at the saddle point. These are saddle points where the variables are just numbers, rather than fields, and so can be found readily. We also compute the next small corrections to the large $p$ limit.

As we review in section \ref{sec:GSigma},  double-scaled SYK has a $G\Sigma $ form, similarly to the finite $p$ SYK model. The $G\Sigma $ action is usually thought of as a gravitational description of the theory. Indeed, in double-scaled SYK, the $G\Sigma $ action reduces to a Liouville action, where the Liouville field corresponds to the 2-point function of SYK. In two dimensional gravity, the Liouville action describes the quantum mechanics of the Weyl factor. This motivates us to study the two dimensional metric in double-scaled SYK induced by the 2-point function.

The parameter $\lambda $ controls the semiclassical limit of the Liouville theory in double-scaled SYK. That is, $\lambda $ is similar to the role of $1/N$ in certain large $N$ theories, or more generally it is analogous to $\hbar$. In the $\lambda  \to 0$ limit, the geometry is classical and very weakly fluctuating. We find that in this classical limit, and at low temperatures, the metric is simply that of (A)dS${}_2$.
As we explain in section \ref{sec:geometry}, the space can be interpreted either as AdS or dS, since the metric differs only by an overall sign.
For the purpose of the presentation solely, we will describe the results in terms of AdS, but there are also reasons to prefer the dS point of view as explained in the text. When we increase the temperature, the space is still rigid AdS, but the boundary is pushed towards the bulk. This is a familiar phenomenon in Jackiw-Teitelboim (JT) gravity.

We then calculate the 1-loop determinant correction to the geometry. In fact, the 1-loop structure is quite intricate. In particular, thinking about the expressions for the SYK observables as integrals over the exponential of an ``action'', this action itself receives a correction at the 1-loop order. At low temperatures, the space is still asymptotically AdS, but is deformed in the bulk. The curvature increases in magnitude as we go into the bulk. It becomes strongly coupled in the interior, where the semiclassical approximation breaks down.

More generally, while the gravitation description of double-scaled SYK at finite $\lambda $ can be more intricate \cite{Berkooz:2022mfk}, we see that at least in an asymptotic expansion around $\lambda =0$ one can still study an ordinary fluctuating bulk metric and spacetime. The fluctuations in the metric, however, become stronger and stronger as we increase $\lambda $. Moreover, we find that the geometrical curvature we study has a simple physical interpretation at finite $\lambda $, in the form of fluctuations in the energy of light operators.

Our discussion is related to the kinematic space description of holography, see e.g., \cite{Czech:2014ppa,Czech:2015qta,Czech:2016xec}. In two dimensions, kinematic space on a particular geometry is the space of oriented geodesics. For a hyperbolic space, this is equivalent to the space of pairs of points on the asymptotic boundary. Our construction is similarly based on pairs of points on the boundary. At leading order, that is, at small $\lambda $ and low temperatures, our metric agrees with the metric on kinematic space, which is the second derivative of the (regularized) length of the geodesic connecting the two boundary points. However, more generally, our geometry differs from that of the kinematic space studies. The holographic kinematic space idea is to relate an emergent bulk geometry to information theoretic quantities on the boundary. More concretely, the metric in kinematic space is related to entanglement entropies on the boundary (or conditional mutual information), according to the Ryu-Takayanagi prescription. In contrast, here we use the 2-point function. Our emergent gravitational construction can be viewed as a different kinematic space notion with the gravitational motivation described above.

Finally, we consider partially entangled thermal states (PETS). These are partially entangled states, generalizing the thermofield double state. We study the entanglement structure in these states in the classical $\lambda  \to 0$ limit.

The outline is as follows.
In section \ref{sec:classical} we calculate the partition function and 2-point function in the classical limit $\lambda  \to 0$ and reproduce the large $p$ results. In section \ref{sec:one_loop} we calculate the next semiclassical approximation including the 1-loop determinant to the results of section \ref{sec:classical}. In section \ref{sec:geometry} we study the geometry induced in the semiclassical approximation. We also calculate it numerically and compare to the semiclassical calculation. Lastly, in section \ref{sec:PETS} we calculate the Renyi entropies of PETS in the classical limit.

\section{Classical limit} \label{sec:classical}

In this section we evaluate the classical limit of double-scaled SYK where $\lambda  \to 0$, keeping only the leading order in the $1/\lambda $ expansion. We consider the partition function and the 2-point function in this section. Higher correlation functions are mentioned in appendix \ref{sec:2n_point_function}.

\subsection{Partition function}

By the partition function of SYK we mean the averaged thermodynamic partition function $\langle \tr e^{-\beta H} \rangle _J$. We use the normalization of the trace where $\tr 1=1$.
The partition function of double-scaled SYK is known \cite{Berkooz:2018qkz,Berkooz:2018jqr}. It is given by an integral over an angular variable $\theta $ that parametrizes the energies.
It was shown in \cite{Berkooz:2018jqr} that the partition function, when we take $\lambda\to0$ and
concentrate on low energies, agrees with the Schwarzian result \cite{Mertens:2017mtv,Lam:2018pvp}. Here, instead,
we consider the regime where $\lambda\to0$, but we keep the energies
finite.

The expression for the partition function, written in terms of $\JMS$, and taking $\lambda $ to be small in the last exponent of the following expression, is
\begin{equation} \label{eq:basic_Z_formula}
Z=\int_{0}^{\pi}\frac{d\theta}{2\pi}(q,e^{\pm2i\theta};q)_{\infty}e^{-\frac{2\beta \JMS}{\lambda}\cos\theta};
\end{equation}
this formula uses the $q$-Pochhammer symbol defined to be
\begin{equation}
(a;q)_{\infty } =\prod _{k=0} ^{\infty } (1-aq^k) .
\end{equation}
In addition, when we use arguments separated by a comma in a $q$-Pochhammer symbol, as well as a $\pm $ sign (as in \eqref{eq:basic_Z_formula}), we mean that there is one such object for every term and every sign. In particular, in \eqref{eq:basic_Z_formula} there is a product of three Pochhammer symbols $(q;q)_{\infty } ( e^{2i\theta } ;q) _{\infty } ( e^{-2i\theta } ;q) _{\infty } $.

For small $\lambda $, at this order, we use the fact that
\begin{equation}
(x;q)_{\infty}\approx\exp\left[-\frac{1}{\lambda}\li_{2}(x)\right]
\end{equation}
where $\li_2(x)$ is the dilogarithm function.

Therefore we can write
\begin{equation} \label{eq:part_func_small_lambda}
Z=\int_{0}^{\pi}\frac{d\theta}{2\pi}(q;q)_{\infty}\exp\left[-\frac{1}{\lambda}\Bigl( \li_{2}(e^{2i\theta})+\li_{2}(e^{-2i\theta})+ {2\beta \JMS \cos(\theta)}\Bigr)\right] .
\end{equation}
In this form, we can manifestly use a saddle point approximation, and we see that $\lambda  \to 0$ corresponds to a classical limit. Using
the definition of the dilogarithm, the saddle point equation is\footnote{In performing the sum over $k$, we had to assume that $0<2\theta<2\pi,$
which is indeed the case.}
\begin{equation}
4\sum_{k=1}^{\infty}\frac{\sin(2\theta k)}{k}+2\beta \JMS\sin(\theta)=2(\pi-2\theta)+2\beta \JMS\sin(\theta)=0.
\end{equation}
To make contact with \cite{Maldacena:2016hyu}, we define
\begin{equation} \label{eq:theta_v_relation}
\theta=\frac{\pi}{2}+\frac{\pi v}{2} .
\end{equation}
We stress that this is merely a definition.
The saddle point equation becomes
\begin{equation} \label{eq:beta_v_rel}
\beta \JMS=\frac{\pi v}{\cos\frac{\pi v}{2}} .
\end{equation}
This equation is familiar from the large $p$ analysis in \cite{Maldacena:2016hyu}. There, this equation is a definition of $v$, that is, a way to parametrize $\beta \JMS$. When $\beta \JMS$ goes from 0 to infinity, $v$ goes from 0 to 1. In contrast, here it is a saddle point equation, relating the energy at the saddle point to the temperature.

The dilogarithms in the exponent of \eqref{eq:part_func_small_lambda} are (we use that $\theta $ goes from 0 to $\pi $)
\begin{equation} \label{eq:Z_sum_of_dilogs}
\li_{2}(e^{2i\theta})+\li_{2}(e^{-2i\theta})=2\sum_{k=1}^{\infty}\frac{\cos(2\theta k)}{k^{2}}=\frac{\pi^{2}}{3}-2\pi\theta+2\theta^{2}=-\frac{\pi^{2}}{6}+\frac{\pi^{2}v^{2}}{2}
.
\end{equation}
Note also that asymptotically
\begin{equation} \label{eq:q_Pochhammer_asymp}
(q;q)_{\infty}=\sqrt{\frac{2\pi }{\lambda } }\exp\left[-\frac{\pi^{2}}{6\lambda}+\frac{\lambda }{24}  \right] .
\end{equation}
We notice that the numerical constant (independent of the temperature) in \eqref{eq:Z_sum_of_dilogs} that contributes as $1/\lambda $ in the exponent of \eqref{eq:part_func_small_lambda} cancels with that in the $q$-Pochhammer symbol \eqref{eq:q_Pochhammer_asymp}. We will treat more carefully the 1-loop determinant in section \ref{sec:one_loop}, but at this order we only care about an overall $\sqrt{\lambda }$ term that we clearly get from it. This term cancels a similar term in \eqref{eq:q_Pochhammer_asymp}, so that there are no overall vanishing or diverging (as $\lambda  \to 0$) numerical coefficients depending on $\lambda $.
 
Therefore, without including subleading $O(\lambda^{0})$ terms,
\begin{equation}
\begin{split} & Z=\exp\left[-\frac{\pi^{2}v^{2}}{2\lambda}-\frac{2\beta \JMS\cos\theta}{\lambda}\right]=\exp\left[-\frac{N}{p^{2}}\frac{\pi^{2}v^{2}}{4}+\frac{N}{p^{2}}\pi v\tan\frac{\pi v}{2}\right]\end{split}
\end{equation}
which is precisely the result in \cite{Maldacena:2016hyu}.\footnote{There is also an obviious factor of $2^{N/2}$ in the partition
function that we do not write, and is absent above because of our normalization $\tr1=1$.}

\subsection{2-point function}

Operators in double-scaled SYK are essentially characterized by the number of fermions that the operators are comprised of. Considering the 2-point function of the same operator, the corresponding parameter is denoted by $\tilde q$. More explicitly, for an operator having $l$ fermions times the number of fermions in the Hamiltonian (which is $p$), $\tilde q=q^l$. Analogously to the relation between $q$ and $\lambda $, we also define
\begin{equation}
\tilde{q}=e^{-\tilde{\lambda}} .
\end{equation}

Here we start by considering the unnormalized 2-point function $\tilde G =\langle \tr e^{-\beta H} \cO (\tau ) \cO (0)\rangle _J$.
We take the Euclidean time variable to go from $0$ to $\beta $. 
The expression for the 2-point function now includes two integrations, and is given by (where again $\lambda$ is
taken to be small)\footnote{In the form of the 2-point function that we use, it does not have definite periodicity properties in $\tau $, but can be made (anti-)periodic by an obvious continuation.}
\begin{equation}
\tilde{G}=\int_{0}^{\pi}\prod_{j=1}^{2}\left\{ \frac{d\theta_{j}}{2\pi}(q,e^{\pm2i\theta_{j}};q)_{\infty}\right\} \exp\left[-\frac{2\tau \JMS\cos\theta_{1}}{\lambda}-\frac{2(\beta-\tau)\JMS\cos\theta_{2}}{\lambda}\right]\frac{(\tilde{q}^{2};q)_{\infty}}{(\tilde{q}e^{i(\pm\theta_{1}\pm\theta_{2})};q)_{\infty}} .
\end{equation}
Just as before, the $\pm $ signs in the last $q$-Pochhammer symbol mean that we have a product of the four corresponding Pochhammer symbols.

As in the partition function, we do not keep terms finite in $\lambda$, so that to order
$1/\lambda$ the 2-point function can be written in the form
\begin{equation}
\tilde{G}=(q;q)_{\infty}^{2}(\tilde{q}^{2};q)_{\infty}\int_{0}^{\pi}\frac{d\theta_{1}d\theta_{2}}{(2\pi)^{2}}
\exp\left[-\frac{f}{\lambda}\right]
\end{equation}
where we define
\begin{equation}
f=\li_{2}\left(e^{\pm2i\theta_{1}}\right)+\li_{2}\left(e^{\pm2i\theta_{2}}\right)-\li_{2}\left(\tilde{q}e^{i(\pm\theta_{1}\pm\theta_{2}}\right)+2\JMS\tau\cos\theta_{1}+2\JMS(\beta-\tau)\cos\theta_{2}
\end{equation}
and by $\pm$ we mean that we should sum over all terms with all possible
signs.

As before, $\lambda  \to 0$ corresponds to a classical limit in which we can evaluate the integral via a saddle point approximation, thinking about $f$ as an ``action''.
The saddle point ``equations of motion''  are\footnote{Note that we work with the branch of the $\log$ function
in which the argument is in the range $(-\pi,\pi)$.} (using $\li_{1}(x)=-\log(1-x)$)
\begin{equation}
\begin{split} & -2i\log\left(1-e^{2i\theta_{1}}\right) +2i\log\left( {1-e^{-2i\theta_{1}}}\right)
\\[2mm] 
& +i\log\bigl(1-e^{-\tilde{\lambda}+i(\theta_{1}\pm\theta_{2})}\bigr)-i\log\bigl({1-e^{-\tilde{\lambda}+i(-\theta_{1}\pm\theta_{2})}}\bigr)-2\JMS\tau\sin\theta_{1}=0
\end{split}
\end{equation}
and the second equation is given by exchanging $\theta_{1}\leftrightarrow\theta_{2}$
and $\tau\leftrightarrow\beta-\tau$.

For general operators, including heavy operators, we can write the saddle point equations as
\begin{equation}
\begin{split}
& 2\theta _1-\pi +\arctan\left[ \frac{\sin(\theta _1+\theta _2)}{e^{\tilde \lambda } -\cos(\theta _1+\theta _2)} \right] +\arctan\left[ \frac{\sin(\theta _1-\theta _2)}{e^{\tilde \lambda } -\cos(\theta _1-\theta _2)} \right] =\JMS \tau \sin\theta _1,\\
& 2\theta _2-\pi +\arctan\left[ \frac{\sin(\theta _1+\theta _2)}{e^{\tilde \lambda } -\cos(\theta _1+\theta _2)} \right] -\arctan\left[ \frac{\sin(\theta _1-\theta _2)}{e^{\tilde \lambda } -\cos(\theta _1-\theta _2)} \right] =\JMS (\beta -\tau) \sin\theta _2 .
\end{split}
\end{equation}

If one is interested in the bulk geometry, we should consider probe operators, that is, operators that are not too heavy in order not to backreact strongly on the geometry. This motivates us to consider small $\tilde \lambda $.

In fact, the saddle point structure here is somewhat subtle. Looking for generic saddle point solutions $\theta _1=\theta ^*_1$ and $\theta _2=\theta ^*_2$ and expanding the equations in small $\tilde \lambda $ would lead to wrong conclusions. Instead, we should consider the same value at leading order.
More precisely, let us work up to order $O(\tilde{\lambda})$ and use the ansatz
\begin{equation}
\begin{split} & \theta_{1}=\theta+\alpha\tilde{\lambda}+O(\tilde{\lambda}^{2}),\\
 & \theta_{2}=\theta-\alpha\tilde{\lambda}+O(\tilde{\lambda}^{2}).
\end{split}
\end{equation}
In fact, this makes sense physically, since for light operators, they should not change the energy by much.

The first saddle point equation then simplifies to
\begin{equation}
i\log\left(-e^{-2i\theta}\right)-2\JMS\tau\sin\theta+2\arctan(2\alpha)+O(\tilde{\lambda})=0.
\end{equation}
So together, the two saddle point equations are now
\begin{equation}
\begin{split} & \begin{cases}
2\theta-\pi-2\JMS\tau\sin\theta+2\arctan(2\alpha)=0,\\
2\theta-\pi-2\JMS(\beta-\tau)\sin\theta-2\arctan(2\alpha)=0.
\end{cases}\end{split}
\end{equation}
Note that $\alpha$ (which was the $O(\tilde{\lambda})$ term) enters
the $O(\tilde{\lambda}^{0})$ equations. This is so because of the singular behavior induced by having the same saddle point value of $\theta _1$ and $\theta _2$ at leading order. Also note that adding $\eta\tilde{\lambda}$
to both $\theta_{1,2}$, will not affect these equations; this can be understood since it can be thought of as shifting $\theta$ at order $\tilde \lambda $. Similarly, one can check that $\tilde{\lambda}^{2}$
terms will not affect these equations, but will enter in the $O(\tilde{\lambda})$
equations.

The sum of these two equations gives the same equation for $\theta$
as in the partition function, and therefore with the definition \eqref{eq:theta_v_relation} of $v$, we get that (\ref{eq:beta_v_rel}) still holds. The other equation
is
\begin{equation}
\pi v+2\arctan(2\alpha)=2\JMS\tau\cos\left(\frac{\pi v}{2}\right).
\end{equation}
We can write the solution to $\alpha $ explicitly as
\begin{equation}
\alpha =- \frac{1}{2} \tan \left[ \frac{\pi v}{2} \left( 1-\frac{2\tau }{\beta } \right) \right] .
\end{equation}
Finally, we should substitute the saddle point values into $f$. A lengthy but straightforward computation gives 
\begin{equation}
\begin{split} 
f & =-\frac{\pi^{2}}{2}+\frac{\pi^{2}v^{2}}{2}-2\pi v\tan\frac{\pi v}{2}+\tilde{\lambda}\Bigg[2-2\log(2\tilde{\lambda})
%\\ & \qquad
 +\log\left(\frac{\cos\bigr(\pi v(\frac{1}{2}-\frac{\tau}{\beta})\bigr)^{2}}{\cos\bigl(\frac{\pi v}{2}\bigr)^{2}}\right)\Bigg]+O(\tilde{\lambda}^{2}).
\end{split}
\end{equation}

At this order, where we do not keep constant $O(\lambda ^0)$ terms, the additional pieces we have are (1) $(\sqrt{\lambda})^{2}$ from
the 1-loop determinant similarly to before, (2) $(q;q)_{\infty}^{2}\sim\frac{1}{\lambda}\exp\left[-\frac{\pi^{2}}{3\lambda}\right]$,
and (3) $\left(\tilde{q}^{2};q\right)_{\infty}\sim\exp\left[-\frac{\pi^{2}}{6\lambda}-\frac{\tilde{\lambda}}{\lambda}\left(-2+2\log(2\tilde{\lambda})\right)\right]$.
We see that again all numerical constants cancel.
We should also normalize the 2-point function by the partition function, $G=\tilde G/Z$, giving altogether at this order
\begin{equation} \label{eq:2pf_leading_order_G}
G=\left[\frac{\cos\frac{\pi v}{2}}{\cos\left(\frac{\pi v}{2}\left(1-\frac{2\tau}{\beta}\right)\right)}\right]^{2\tilde{\lambda}/\lambda}.
\end{equation}
When we recall the $G\Sigma $ formalism in section \ref{sec:GSigma}, we will have a corresponding variable $g$, in terms of which
\begin{equation} \label{eq:2pf_leading_order}
e^g=\left[\frac{\cos\frac{\pi v}{2}}{\cos\left(\frac{\pi v}{2}\left(1-\frac{2\tau}{\beta}\right)\right)}\right]^{2}.
\end{equation}
This agrees exactly with \cite{Maldacena:2016hyu}.

\section{1-loop determinant} \label{sec:one_loop}

In this section we go beyond the classical approximation. In order to account for the leading order correction in small $\lambda $ to the classical result, we consider the 1-loop determinant. In fact, it is important that in our model the integrand itself, which can be thought of as an action, receives a correction as well, which should be taken into account. Moreover, in order to get the right answer, the saddle point structure gets corrected in a way that affects the result in a nontrivial manner. We explain this below.

\subsection{Partition function}

Let us start with the partition function where the situation is simpler to analyze.
There are two sources of corrections to the ``action'' itself. One is from the asymptotic expansion of the $q$-Pochhammer symbol which to order $O(\lambda ^0)$ is
\begin{equation}
(x;q)_{\infty}=\exp\left[-\frac{1}{\lambda}\li_{2}(x)+\frac{1}{2} \log(1-x) + O(\lambda)\right] .
\end{equation}
In addition, we should take into account the expansion of the energies themselves.\footnote{In terms of the chords coupling $\cJ$, the energy is given by $E(\theta )=\frac{2\cJ \cos \theta }{\sqrt{1-q}} $, see \cite{Berkooz:2018jqr}.} Together we get that we need to evaluate
\begin{equation}
\begin{split}
Z&= \int \frac{d\theta }{2\pi } (q;q)_{\infty } \exp \Bigg[-\frac{2\beta \JMS \cos \theta }{\lambda } -\frac{\beta \JMS\cos\theta }{2} -\frac{1}{\lambda } \li_2\left( e^{2i\theta } \right) -\frac{1}{\lambda } \li_2\left( e^{-2i\theta } \right) +\\
& \qquad +\frac{1}{2} \log\left( \left( 1-e^{2i\theta } \right) \left( 1-e^{-2i\theta } \right) \right) \Bigg].
\end{split}
\end{equation}
For the 1-loop determinant, we should consider the second derivative of the coefficient of $(-1/\lambda) $ which is $4-2\beta \JMS \cos\theta$.

Plugging in the saddle point value in the integrand and including the 1-loop determinant contribution to the saddle point evaluation, we find at this order the result
\begin{equation}
Z = \frac{\cos \frac{\pi v}{2} }{\sqrt{1+\frac{\pi v}{2} \tan \frac{\pi v}{2} }} 
\exp\left[-\frac{\pi^{2}v^{2}}{2\lambda }+\frac{2\pi v}{\lambda }\tan\frac{\pi v}{2} + \frac{\pi v}{2} \tan \frac{\pi v}{2}  \right].
\end{equation}
We can write the free energy as
\begin{equation}
\beta F = \frac{N}{p^2} \frac{\pi ^2v^2}{4} -\frac{N}{p^2} \pi v \tan \frac{\pi v}{2} -\frac{\pi v}{2} \tan \frac{\pi v}{2} - \log \cos \frac{\pi v}{2} +\frac{1}{2} \log\left( 1+\frac{\pi v}{2} \tan \frac{\pi v}{2} \right) .
\end{equation}
The three additional last terms serve as small corrections to the leading order result found in \cite{Maldacena:2016hyu}.

\subsection{2-point function}

The evaluation of the 2-point function starts similarly to the partition function, but there is going to be an important difference. We first of all write the corrections to the integrand coming from the energies and the expansion of the matrix elements of the operators. We denote the subleading, order $\lambda ^0$, part of the integrand by $h(\theta _1,\theta _2)$. The unnormalized 2-point function is then written as
\begin{equation}
\tilde{G}=(q;q)_{\infty}^{2}(\tilde{q}^{2};q)_{\infty}\int_{0}^{\pi}d\theta_{1}d\theta_{2}
\, h(\theta _1,\theta _2) \exp\left[-\frac{f}{\lambda}\right]
\end{equation}
where
\begin{equation}
\begin{split}
& f=\li_{2}\left(e^{\pm2i\theta_{1}}\right)+\li_{2}\left(e^{\pm2i\theta_{2}}\right)-\li_{2}\left(\tilde{q}e^{i(\pm\theta_{1}\pm\theta_{2}}\right)+2\tau\JMS\cos\theta_{1}+2(\beta-\tau)\JMS\cos\theta_{2}\\
& h = \frac{1}{4\pi ^2}  \exp \Bigg[ -\frac{\tau \JMS\cos\theta_{1}}{2}-\frac{(\beta-\tau)\JMS\cos\theta_{2}}{2}
+ \frac{1}{2} \log \left( 1-e^{\pm 2i\theta _1}\right)  + \frac{1}{2} \log \left( 1-e^{\pm 2i\theta _2} \right) 
-\\
& \qquad - \frac{1}{2} \log \left( 1-\tilde q e^{i(\pm \theta _1 \pm \theta _2)}\right)  \Bigg]
\end{split}
\end{equation}
and again by $\pm$ we mean that we sum over all terms with all possible
signs.

Before we evaluate the 1-loop determinant, we note the following important point. To the order we work in, both in the 1-loop determinant and the integrand correction $h$, we have to go to higher order in the saddle point values. The reason is similar to before, since some information (but not all) of higher orders in the saddle point values enters in lower order when substituting back in the action and the 1-loop determinant.

The ansatz that we should now use for the saddle point is
\begin{equation}
\begin{split}
\theta _1 &= \theta +\alpha \tilde \lambda +\eta  \tilde \lambda +\gamma \tilde \lambda ^2,\\
\theta _2 &= \theta -\alpha \tilde \lambda +\eta \tilde \lambda -\gamma \tilde \lambda ^2.
\end{split}
\end{equation}
Note that at the leading classical level in section \ref{sec:classical}, it did not matter what $\eta $ is, that is, it did not enter in the saddle point equations. Now, it is important to determine $\eta $. For the second order term (the coefficient of $\tilde \lambda ^2$), similarly to before, only the difference of the coefficients is important, just as happened with $\alpha $ at the previous order. The values of $\alpha $ and $\theta $ are the same as before, while the $O(\tilde \lambda )$ order in the saddle point equations fixes
\begin{equation}
\begin{split}
\eta &= -\frac{2+\pi v\left( 1-\frac{2\tau }{\beta } \right) \tan\left[ \frac{\pi v}{2 } \left( 1-\frac{2\tau }{\beta } \right) \right] }{2\pi v+4\cot \frac{\pi v}{2} } ,\\
\gamma &= \frac{\tan\left( \frac{\pi v}{2} \right) }{4\cos\left[ \frac{\pi v}{2} \left( 1-\frac{2\tau }{\beta } \right) \right] ^2 \left( \pi v+2\cot \frac{\pi v}{2} \right) } \Bigg[ -\pi v\left( 1-\frac{2\tau }{\beta } \right) +\\
& \qquad \qquad \qquad + 2\left( \pi v\left(1-\frac{\tau }{\beta } \right) +\cot \frac{\pi v}{2} \right) \left( \frac{\pi v\tau }{\beta } +\cot \frac{\pi v}{2} \right) \tan\left( \frac{\pi v}{2} \left( 1-\frac{2\tau }{\beta } \right) \right) \Bigg] .
\end{split}
\end{equation}
Importantly, these corrections do not affect the $O(\tilde \lambda )$ order of $f$ and this is why this issue was not important when considering the partition function in the previous subsection.

Working up to order $O(\tilde \lambda )$ in the 1-loop determinant corrections, the 2-point function becomes
\begin{equation}
\begin{split}
& \tilde G = (q;q)_{\infty } ^2 (\tilde q^2;q)_{\infty } \cdot \\
& \qquad \frac{1}{4\pi ^2} \exp \Bigg[ \log 2 + \log \cos \frac{\pi v}{2} -\log \tilde \lambda +\log \cos \left( \frac{\pi v}{2} \left( 1-\frac{2\tau }{\beta } \right) \right) +\frac{\pi v}{2} \tan \frac{\pi v}{2} +h_1\tilde \lambda \Bigg] \cdot \\[2mm]
& \qquad \exp \Bigg[ \frac{\pi ^2}{2\lambda } -\frac{\pi ^2v^2}{2\lambda } +\frac{2\pi v}{\lambda } \tan \frac{\pi v}{2} - \frac{\tilde \lambda }{\lambda } \left( 2-2\log(2 \tilde \lambda)   +   \log\left( \frac{ \cos \bigl( \frac{\pi v}{2} \bigl( 1-\frac{2\tau }{\beta } \bigr)\bigr)^2}{\cos\bigl( \frac{\pi v}{2}\bigr)^2}\right)\right)  \Bigg] \cdot\\ &\qquad
\frac{2\pi \lambda \sqrt{\tilde \lambda }}{2\cos\left( \frac{\pi v}{2} \left( 1-\frac{2\tau }{\beta } \right) \right) \sqrt{2+\pi v\tan \frac{\pi v}{2} }}  \left( 1-\frac{\tilde \lambda }{2} \frac{f''_0}{4\cos\left( \frac{\pi v}{2} \left( 1-\frac{2\tau }{\beta } \right) \right) ^2\left( 2+\pi v\tan \frac{\pi v}{2} \right) } \right) 
\end{split}
\end{equation}
where we denoted the subleading corrections
\begin{equation}
\begin{split}
& h_1 = \frac{1}{8\left( 2+\pi v\tan\frac{\pi v}{2} \right)} \cos\left( \frac{\pi v}{2} \right) ^{-2} \cos\Bigl( \frac{\pi v}{2}\Bigl( 1-\frac{2\tau }{\beta } \Bigr) \Bigr) ^{-2} \cdot \\[1mm]
& \qquad \Bigg[ 8+2\pi ^2v^2 \frac{\tau }{\beta } -2\pi ^2v^2\frac{\tau ^2}{\beta ^2} + \left( 4-2\pi ^2v^2 \frac{\tau }{\beta } +2 \pi ^2v^2 \frac{\tau ^2}{\beta ^2} \right) \left[ \cos(\pi v)+ \cos\Bigl({\pi v}\Bigl( 1-\frac{2\tau }{\beta } \Bigr) \Bigr) \right]  +\\[1mm]
& \qquad +\pi ^2v^2 \frac{\tau }{\beta }\Bigl( 1-\frac{\tau }{\beta } \Bigr)  \left[  \cos\Bigl( 2{\pi v}\Bigl( 1-\frac{\tau }{\beta } \Bigr) \Bigr) +\cos \Bigl( 2\pi v \frac{\tau }{\beta } \Bigr) \right] +3\pi v \sin(\pi v)+\\[1mm]
& \qquad + \pi v\Bigl( 1-\frac{2\tau }{\beta } \Bigr) \sin\Bigl( {\pi v}\Bigl( 1-\frac{2\tau }{\beta } \Bigr) \Bigr) -\frac{\pi v\tau }{\beta } \sin\Bigl( 2{\pi v}\Bigl( 1-\frac{\tau }{\beta } \Bigr) \Bigr) -\pi v\bigl( 1-\frac{\tau }{\beta } \bigr) \sin \frac{2\pi v\tau }{\beta } \Bigg]
\end{split}
\end{equation}
and
\begin{equation}
\begin{split}
f''_0  =& -\frac{4\pi v \cos \left[ \frac{\pi v}{2} \left( 1-\frac{2\tau }{\beta } \right) \right] ^2}{\pi v+2\cot \frac{\pi v}{2} }+ \frac{12 \left( \cos \frac{\pi v}{2} +\pi v\left( 1-\frac{\tau }{\beta } \right) \sin \frac{\pi v}{2} \right) \left( \cos \frac{\pi v}{2} +\frac{\pi v\tau }{\beta } \sin \frac{\pi v}{2} \right)}{{\cos\left( \frac{\pi v}{2} \right) ^2  }}- \\[2mm]
&  - {\cos \left[\mbox{\large $ \frac{\pi v}{2} \bigl( 1-\frac{2\tau }{\beta } \bigr)$} \right] ^2}\Bigg[ \frac{4\left[ \left( 1-\frac{\pi ^2v^2\tau }{\beta } +\frac{\pi ^2v^2\tau ^2}{\beta ^2} \right) \cos (\pi v)+\pi v \left( \frac{\pi v\tau (\beta -\tau )}{\beta ^2} +\sin(\pi v) \right)  \right]}{{\cos\left( \frac{\pi v}{2} \right) ^2  } } -\\[2mm]
&\; -\; \frac{\pi v}2\Bigl( 1-\mbox{\large $\frac{2\tau }{\beta }$} \Bigr)\; \frac{\cos \frac{\pi v}{2} +3 \cos \frac{3\pi v}{2} -4\pi v \sin \left( \frac{\pi v}{2} \right) ^3 }{{\cos\left( \frac{\pi v}{2} \right) ^3  } \left( 1+\frac{\pi v}{2}\tan \frac{\pi v}{2} \right)}\tan \biggl( \frac{\pi v}{2} \Bigl( 1-\frac{2\tau }{\beta } \Bigr)\biggr) \,  \Bigg] .
\end{split}
\end{equation}

The normalized 2-point function is then
\begin{equation} \label{eq:2pf_1_loop}
G = \left[ \frac{\cos \left( \frac{\pi v}{2}\right) ^2 }{\cos\left( \frac{\pi v}{2} \left( 1-\frac{2\tau }{\beta } \right) \right)^2 } \left( 1+G_1 \lambda \right) \right]^{\tilde \lambda /\lambda }
\end{equation}
where
\begin{equation}
G_1 = -\frac{1}{2} +h_1-\frac{1}{2} \cdot \frac{f''_0}{4\cos\left( \frac{\pi v}{2} \left( 1-\frac{2\tau }{\beta } \right) \right) ^2\left( 2+\pi v\tan \frac{\pi v}{2} \right) } .
\end{equation}

\section{Induced geometry in the bulk} \label{sec:geometry}

In this section we use the results of sections \ref{sec:classical} and \ref{sec:one_loop} in order to study an emergent semiclassical bulk geometry.

\subsection{The $G\Sigma $ formalism: Liouville form of double-scaled SYK} \label{sec:GSigma}

We briefly recall the $G\Sigma $ description of double-scaled SYK.
The partition function of SYK can be written in a path integral form using the corresponding Lagrangian
\begin{equation}
Z=\int D\psi\:\exp\left\{ -\int d\tau\left[\frac{1}{4} \sum_{i}\psi_{i}\partial_{\tau}\psi_{i}+i^{p/2}\sum J_{i_{1}\cdots i_{p}}\psi_{i_{1}}\cdots\psi_{i_{p}}\right]\right\} 
\end{equation}
where $\tau $ is the Euclidean time and $J_{i_{1}\cdots i_{p}}$ are the random couplings with gaussian variance \eqref{gaussianj}.
At large $N$, annealed disorder is the same as quenched disorder in the theory, so we can consider the disorder averaged partition function 
\begin{equation}
\label{eq:zaverage}
\langle Z\rangle_{J} =\int D\psi\,\exp\left\{ -\int d\tau\left[\frac{1}{4} \sum_{i}\psi_{i}\partial_{\tau}\psi_{i}\right]+\frac{\JMS^{2}}{4p^{2}N^{p-1}}\int d\tau_1 d\tau_2\,\biggl(\sum_{i}\psi_{i}(\tau_1)\psi_{i}(\tau_2)\biggr)^{p}\right\} .
\end{equation}
We now introduce the dynamical mean field variables $G(\tau_1 ,\tau_2)$ and $\Sigma (\tau_1 ,\tau_2)$, both antisymmetric with respect to $\tau _1 \leftrightarrow \tau _2$, by inserting the functional delta function imposing the identification
\begin{equation} \label{eq:inserting_delta}
G(\tau_1,\tau_2) = \frac 1 N \sum_{i}\psi_{i}(\tau_1)\psi_{i}(\tau_2)
\end{equation}  
leading to the intermediate identity
\begin{equation}
\begin{split}
\langle Z\rangle_{J} = 
\int DGD\Sigma D\psi & \exp\left\{ -\int d\tau\left[\frac{1}{4}\sum_{i}\psi_{i}\partial_{\tau}\psi_{i}\right]+N\frac{\JMS^{2}}{4p^{2}}\int d\tau_1 d\tau_2\,G(\tau_1,\tau_2)^{p}\right\} \\
 & \exp\left\{ -\frac{N}{2}\int d\tau_1 d\tau_2\,\Sigma(\tau_1,\tau_2)\biggl(G(\tau_1,\tau_2) -\frac{1}{N}\sum_{i}\psi_{i}(\tau_1)\psi_{i}(\tau_2)\biggr)\right\} .
\end{split}
\end{equation}
Performing the integral over $\Sigma $ along the imaginary axis produces the delta function imposing \eqref{eq:inserting_delta}, which can then be used to eliminate the $G$-integral. This reproduces the original expression \eqref{eq:zaverage}. If instead we integrate out the fermions first, we obtain the following $G\Sigma$ partition function and effective action
\begin{equation}
\langle Z\rangle_{J} =\int DGD\Sigma\,e^{-I[G,\Sigma]}
\end{equation}
\begin{equation}
I=-\frac{N}{2}\log\det\bigl(\delta(\tau_{12})\partial_{\tau_2}-2\Sigma(\tau_1,\tau_2)\bigr)+\frac{N}{2}\int d\tau_1 d\tau_2\left(\Sigma(\tau_1,\tau_2) G(\tau_1,\tau_2)-\frac{\JMS^{2}}{2p^{2}}G(\tau_1,\tau_2)^{p}\right)
\end{equation}
with $\tau_{12} = \tau_1-\tau_2$.
This $G\Sigma$ effective theory is non-local in time. However, in the double scaling limit, it becomes a bi-local theory in $\tau$, or equivalently, a local theory on the `kinematic space' labeled by pairs of time instances $(\tau_1,\tau_2)$. Following \cite{Cotler:2016fpe}, in the double-scaling limit we make the following Ansatz
\begin{equation} \label{eq:GSigma_to_Liouville_vars}
\Sigma(\tau_1,\tau_2)=\frac{\sigma(\tau_1,\tau_2)}{p},\qquad G(\tau_1,\tau_2)=\sign(\tau_{12}) \left(1+\frac{g(\tau_1,\tau_2)}{p}\right)
\end{equation}
where now $g(\tau _1,\tau _2)$ is symmetric while $\sigma (\tau _1,\tau _2)$ is still antisymmetric under $\tau _1 \leftrightarrow \tau _2$.
Expanding in large $p$, the linear term in $\sigma $ cancels between the two contributions to the action, and we remain with (using the fact that $p$ is even, and dropping an additive constant from the determinant term)
\begin{equation}
\begin{split} & I=\frac{N}{4p^{2}}\Bigg[\int d\tau_{1}d\tau_{2}d\tau_{3}d\tau_{4}\sign(\tau_{12})\sigma(\tau_{2},\tau_{3})\sign(\tau_{34})\sigma(\tau_{4},\tau_{1})+\\
 & \qquad+2\int d\tau_{1}d\tau_{2}\left(\sign(\tau_{12})\sigma(\tau_{1},\tau_{2})g(\tau_{1},\tau_{2})-\JMS^{2}e^{g(\tau_{1},\tau_{2})}+O(1/p)\right)\Bigg].
\end{split}
\end{equation}
Performing the Gaussian integral over $\sigma$, using the fact that $g(\tau _1,\tau _2)$ vanishes at $\tau _1=\tau _2$ since the UV boundary condition is $G \to \sign(\tau _1-\tau _2)$ as $\tau _1 \to \tau _2$, we arrive at \cite{Cotler:2016fpe}
\begin{equation}
I=\frac{N}{16p^{2}}\int d\tau_1 d\tau_2\left[\partial_{\tau_1}g(\tau_1,\tau_2)\partial_{\tau_2}g(\tau_1,\tau_2)-4\JMS^{2}e^{g(\tau_{1},\tau_{2})}\right].
\end{equation}
This is a Liouville action.

The expectation value of the correlation function depends only on $\tau _1-\tau _2$. For this reason, and as explained more below, we find it more convenient to use the variables
\begin{equation}
\begin{split}
\tau &= \tau _1-\tau _2,\\
\nu &= \tau _1+\tau _2.
\end{split}
\end{equation}
Expressed in terms of them,
\begin{equation} \label{eq:Liouville_ds_SYK}
I=\frac{N}{16p^{2}}\int d\tau d\nu\left[ -\frac{1}{2} (\partial _{\tau } g)^2 + \frac{1}{2} (\partial _{\nu } g)^2 -2\JMS^{2}e^{g(\tau,\nu )}\right].
\end{equation}
The path integral does not seem to converge in the naive integration contour. We can deform it at large values (essentially by a phase $e^{i\pi /4} $) in order to improve its convergence properties.

\subsection{The metric}

As we saw in section \ref{sec:GSigma}, double-scaled SYK has a Liouville form \eqref{eq:Liouville_ds_SYK}. When quantizing gravity in two dimensions, Liouville theory describes the quantum mechanics of the Weyl factor with respect to the naive metric appearing in the kinetic term. Therefore, the form \eqref{eq:Liouville_ds_SYK} motivates us to consider the metric
\begin{equation} \label{eq:metric_from_Liouville_dS}
ds^2 = e^g \left( -d\tau ^2+d\nu ^2\right)
\end{equation}
which upon imposing the Liouville equations of motion for $g$ takes the form of a two-dimensional de Sitter space-time, or
\begin{equation} \label{eq:metric_from_Liouville}
ds^2 = e^g \left( -d\nu ^2+d\tau ^2\right) .
\end{equation}
which upon imposing the Liouville equations of motion for $g$ takes the form of a two-dimensional Anti-de Sitter space-time.

In the discussions of kinematic space \cite{Czech:2015qta}, one encounters a similar ambiguity. In that case, kinematic space naturally comes with a volume form, which can be used to express the length of a curve in the bulk. In this mapping from the volume form to a metric, one usually chooses a de Sitter space signature. Indeed, two dimensional dS and AdS have the same isometries and same topology (for wrapped AdS). In our discussion, one would expect that the lorentzian Liouville action \eqref{eq:Liouville_ds_SYK} may indicate a preference between one or the other signature. However, this is not so clear-cut, since the action is not positive definite, and the contour deformation discussed above can result in different signatures. For the sole purpose of presentation, we will choose the second option \eqref{eq:metric_from_Liouville} which will give rise to an Anti-de Sitter geometry in the classical limit. However, the analogous discussion for \eqref{eq:metric_from_Liouville_dS} should be clear and equally worth exploring.

Importantly, we see that $\lambda $
controls the semiclassical limit in the Liouville theory. For very small $\lambda $, the theory is semiclassical and has a well defined metric in the sense above. As we increase $\lambda $, the fluctuations become larger. However, we can still study the expectation value of the geometry, remembering that the fluctuations become large.

In the expectation value, $g=g(\tau )$ depends only on $\tau $ and not on $\nu $. The origin of this is the time translation invariance. In the bulk, this statement corresponds to the presence of a timelike Killing vector (in the convention \eqref{eq:metric_from_Liouville}).

Let us start with the leading order solution \eqref{eq:2pf_leading_order_G}. When expressed in terms of the Liouville variable $g$, which is defined through \eqref{eq:GSigma_to_Liouville_vars}, we have \eqref{eq:2pf_leading_order}. Moreover, let us first consider low temperatures, such that $v \to 1$. The metric \eqref{eq:metric_from_Liouville} is then
\begin{equation} \label{eq:classical_metric_low_temp_fixed_tau}
ds^2 = \frac{-d\nu ^2+d\tau ^2}{(1+\JMS \tau )^2} .
\end{equation}
The space we obtain is Lorentzian AdS${}_2$. Note that here, in \eqref{eq:classical_metric_low_temp_fixed_tau}, we keep $\tau $ fixed, so we do not cover the entire space. We can instead rescale $\tau$, and define $\sigma = \frac{\pi \tau }{\beta } $, as well as rescale $\nu $ such that $\nu =\frac{\beta \tilde \nu }{\pi } $. In this case, \eqref{eq:metric_from_Liouville} and \eqref{eq:2pf_leading_order} become more directly
\begin{equation}
ds^2 = \JMS^{-2} \frac{-d\tilde \nu ^2+d\sigma ^2}{\sin(\sigma )^2} .
\end{equation}
This is the metric of global Lorentzian AdS${}_2$ where $\sigma  \in (0,\pi )$.

Indeed, we can also calculate the Ricci scalar from \eqref{eq:metric_from_Liouville} using
\begin{equation}
R = -e^{-g} \nabla ^2 g
\end{equation}
and find
\begin{equation}
R=-2\JMS^2.
\end{equation}
That is, we get constant negative curvature, where $1/\JMS$ sets the radius of AdS.\footnote{If we were to use \eqref{eq:metric_from_Liouville_dS}, we would have found similarly a constant positive curvature.} At this order we have more isometries than in the general case.

We may also consider finite $v<1$, that is, increase the temperature.
Most directly, the metric takes the form
\begin{equation}
ds^2 = \frac{\cos\left( \frac{\pi v}{2} \right) ^2}{\cos \left( \frac{\pi v}{2} \left( 1-\frac{2\tau }{\beta } \right) \right) ^2} ( -d\nu ^2+d\tau ^2 ) .
\end{equation}
This is still a patch in AdS${}_2$; one can verify that we still have the same Ricci scalar. 
The range of the coordinate $\tau $ between the surfaces where the metric blows up is now
$\Delta \tau =\frac{\beta }{v} $. At small temperatures where $v \to 1$ we have instead a range of $\beta $.
Therefore, if we keep the range of $\tau $ to be $(0,\beta )$, we see that for $v<1$ we are left with a smaller patch of AdS. That is, the boundaries get pushed into the bulk. This is a familiar effect in JT gravity when we increase the temperature, or reduce $\beta $ \cite{Maldacena:2016upp,Kitaev:2018wpr}.

\subsection{Semiclassical geometry}

Now let us consider the leading correction to the metric coming from the 1-loop determinant. We will mostly be interested directly in the Ricci scalar.\footnote{We should stress that we discuss here the Ricci scalar that corresponds to the expectation value of the metric, which in principle could differ from the actual expectation value of the curvature. We thank J.~Maldacena for this comment.} Using the expression \eqref{eq:2pf_1_loop}, we get
\begin{equation}
R = -2\JMS^2(1-G_1\lambda ) - \frac{\cos\left( \frac{\pi v}{2} \left( 1-\frac{2\tau }{\beta } \right) \right) ^2}{\cos\left( \frac{\pi v}{2} \right) ^2} \nabla ^2 G_1 \lambda \equiv-2\JMS^2+\Delta R\cdot \lambda .
\end{equation}
We defined $\Delta R$ to be the change from the background AdS.

Using the calculation of $G_1$ in section \ref{sec:one_loop}, we obtain the following explicit analytic result
\begin{eqnarray} \label{eq:Ricci_1_loop}
&& \nonumber \frac{\Delta R}{\JMS^2} =\frac{\cos \left( \frac{\pi v}{2} \right)^{-2} \cos \bigl( \frac{\pi v}{2} \bigl( 1-\frac{2\tau }{\beta } \bigr) \bigr) ^{-2} }{4\left( 2+\pi v \tan \frac{\pi v}{2} \right) } 
\Bigg\{
2\left( 2-\frac{\pi ^2v^2\tau }{\beta } +\frac{\pi ^2v^2\tau ^2}{\beta ^2} \right) \cos(\pi v)-\\
&& \nonumber - 2\left(2 +\frac{\pi ^2v^2\tau }{\beta } -\frac{\pi ^2v^2\tau ^2}{\beta ^2} \right) \cos\Bigl( \pi v\bigl( 1-\frac{2\tau }{\beta } \bigr) \Bigr) + \pi v\Bigg[
\frac{2\pi v(\beta -\tau )\tau }{\beta ^2} 
+2\sin(\pi v)\, +\\
& & \nonumber +\frac{\pi v(\beta -\tau )\tau }{\beta ^2} \cos\Bigl( \pi v\bigl( 1-\frac{\tau }{\beta } \bigr) \Bigr)
+\frac{\pi v(\beta -\tau )\tau }{\beta ^2} \cos \frac{2\pi v\tau }{\beta } 
-2\bigl( 1-\frac{2\tau }{\beta } \bigr) \sin\Bigl( \pi v\bigl( 1-\frac{2\tau }{\beta } \bigr) \Bigr) -\\
& & - \frac{2\tau }{\beta } \sin\Bigl( \pi v\bigl( 1-\frac{\tau }{\beta } \bigr) \Bigr) -2\bigl( 1-\frac{\tau }{\beta } \bigr) \sin \frac{2\pi v\tau }{\beta } \Bigg] \Bigg\} .
\end{eqnarray}

\begin{figure}[h]
\centering
\includegraphics[width=0.5\textwidth]{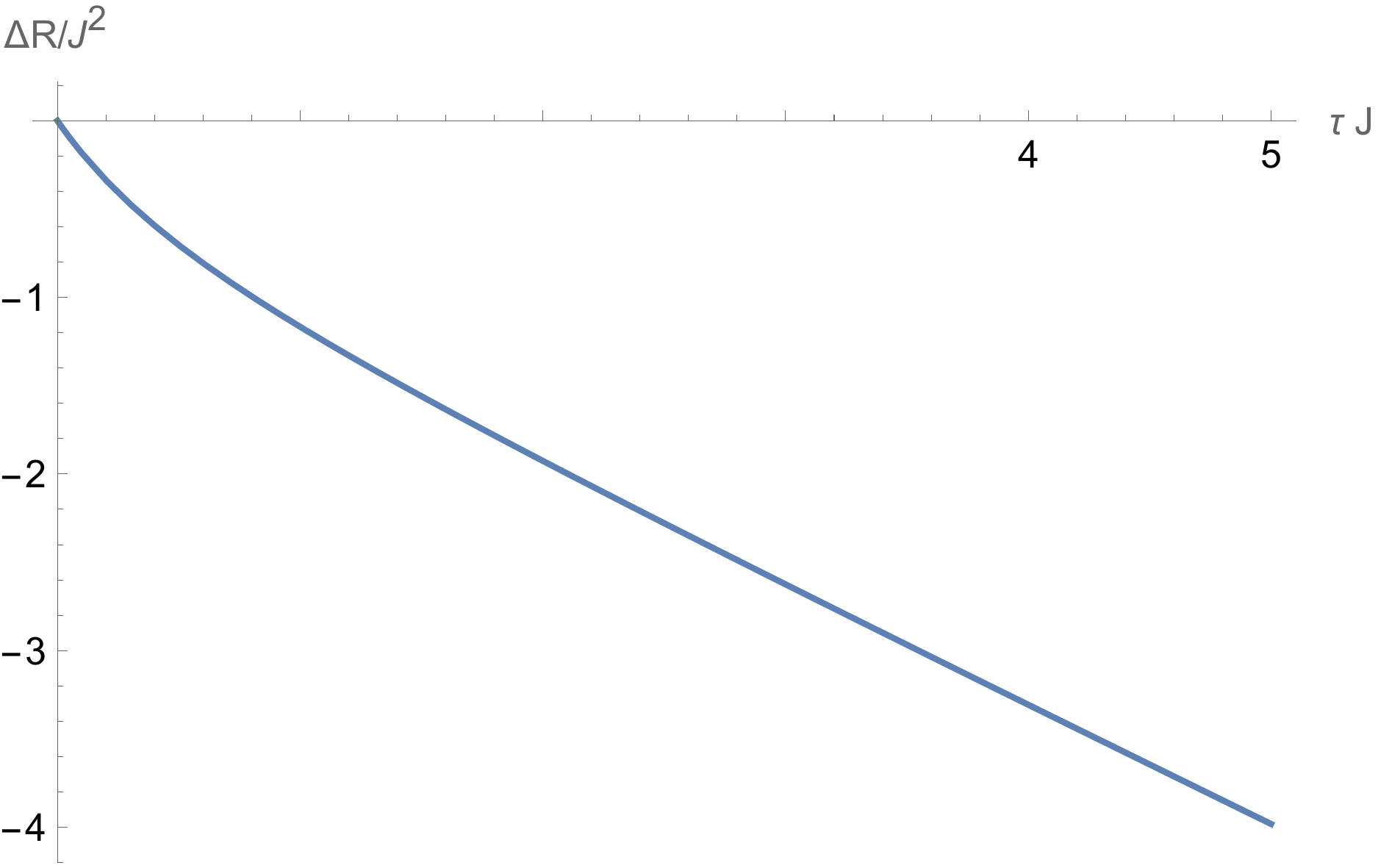}
\caption{The correction to the curvature at small temperatures.}
\label{fig:R_correction_small_T}
\end{figure}

Let us examine the correction to the curvature first for low temperatures. Taking $v \to 1$, we find
\begin{equation}
\Delta R=-\frac{2\JMS \tau (3+3\JMS \tau +\JMS^2 \tau ^2)}{3(1+\JMS \tau )^2} \JMS^2.
\end{equation}
A plot of this is shown in Fig.\ \ref{fig:R_correction_small_T}. At the asymptotic boundary, we see that there is no correction, and we still have an asymptotically AdS space. However, the curvature becomes more and more negative as we go into the bulk (asymptotically linearly with $\tau $). This means that the semiclassical approximation breaks down as we go deep into the bulk, where the theory is strongly coupled.

\begin{figure}[h]
\centering
\includegraphics[width=0.5\textwidth]{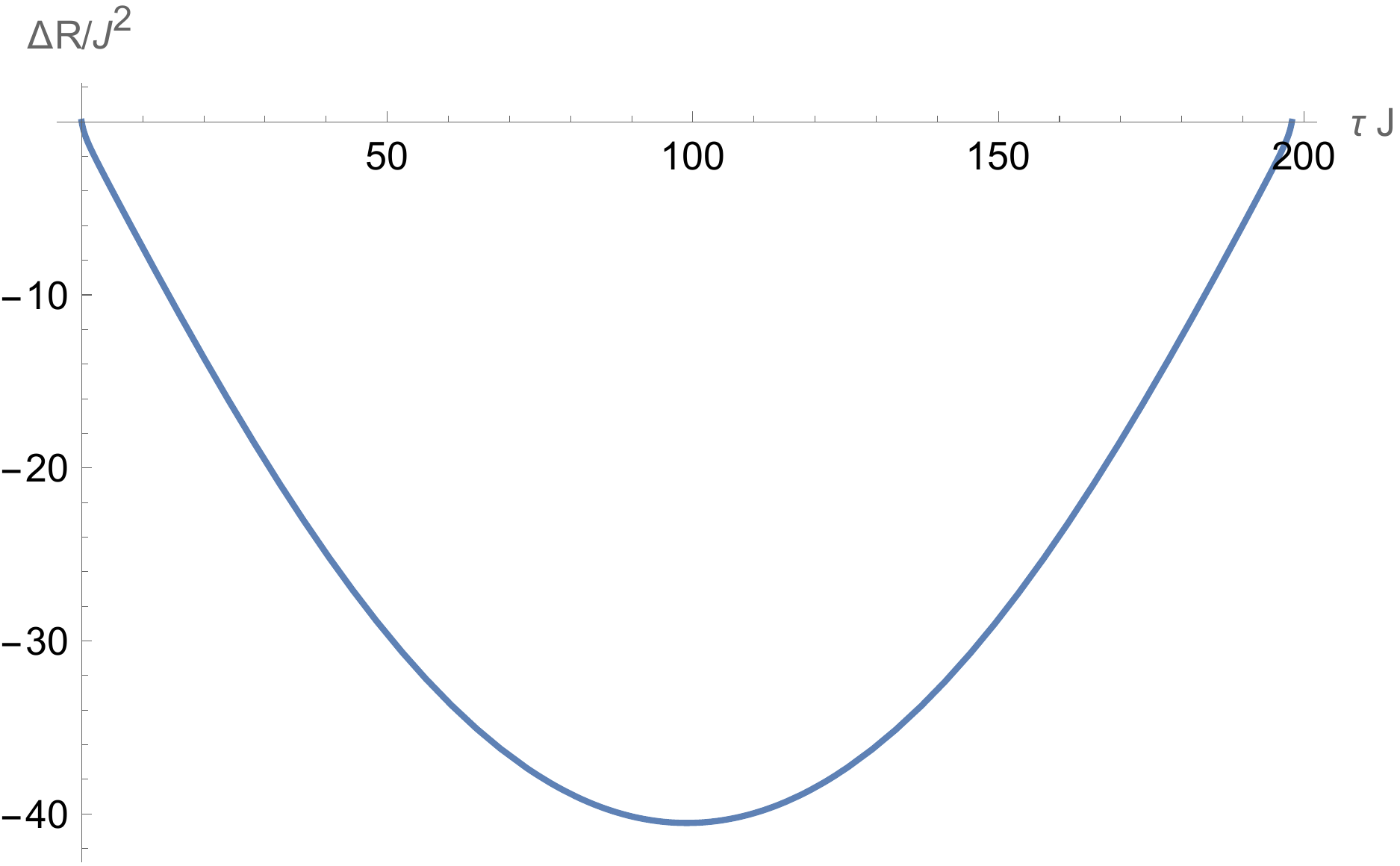}
\caption{Numerical plot of $\Delta R$ at $v=0.99$ for $\tau \in (0,\beta )$.}
\label{fig:R_correction_v_099}
\end{figure}

Since the metric we find deforms the full Lorentzian AdS space (at low temperatures), the curvature is symmetric with respect to $\tau  \to \beta -\tau $. Indeed, we show numerically the correction to the curvature for $\tau  \in (0,\beta )$ and $v=0.99$ in Fig.\ \ref{fig:R_correction_v_099}. When $\tau $ is large, it is indeed linear, until we are getting to the region well deep in the bulk.

\begin{figure}[h]
\centering
\includegraphics[width=0.5\textwidth]{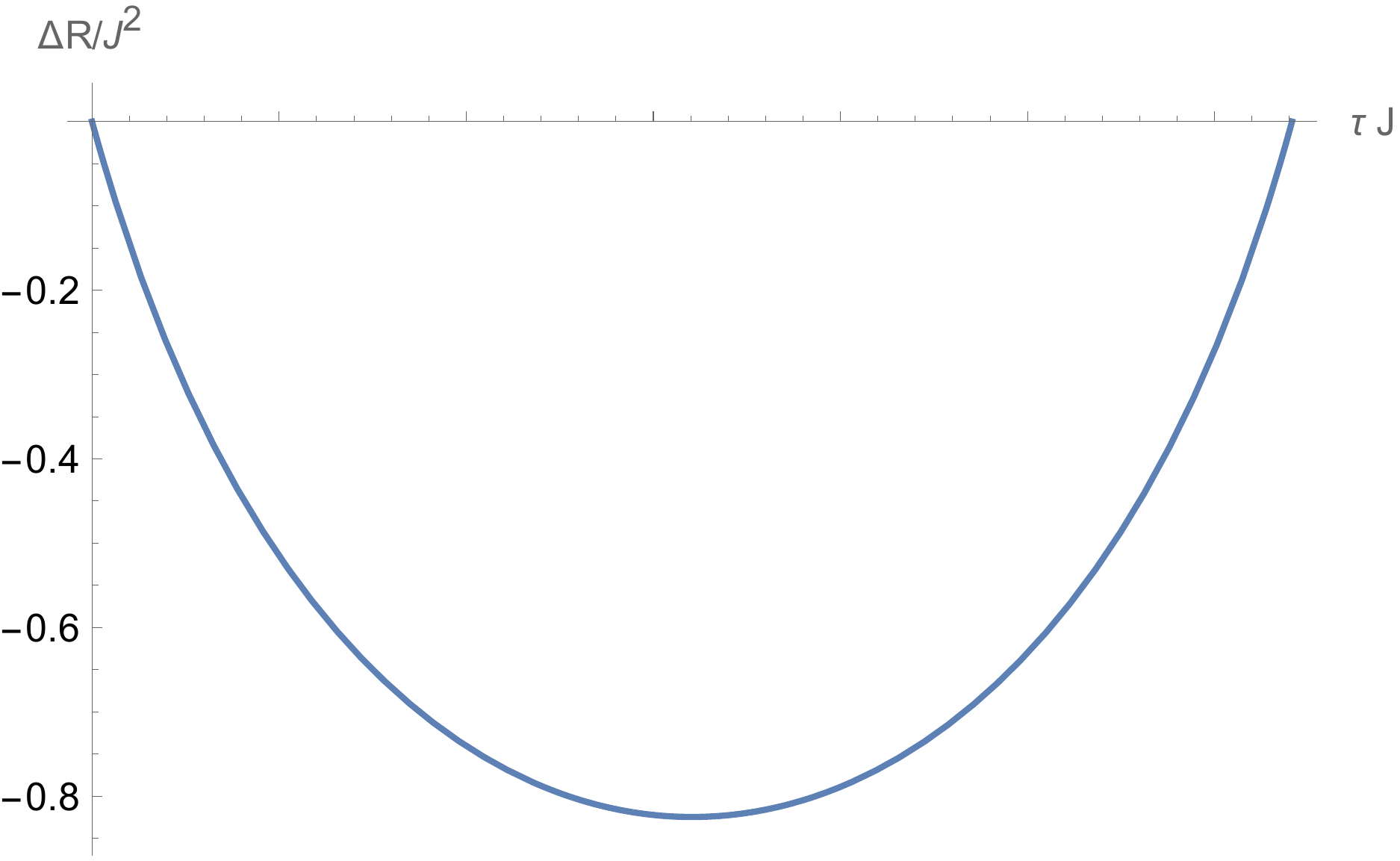}
\caption{Numerical plot of $\Delta R$ at $v=0.6$.}
\label{fig:R_correction_v_06}
\end{figure}

As we increase the temperature further, the curve becomes more smooth. An example is shown in Fig.\ \ref{fig:R_correction_v_06}. Let us write the asymptotic form of the correction for very large temperature, or $v \to 0$. For that we should rescale $\tau $, since $\beta  \to 0$. Defining $\tilde \tau =\tau /\beta $, we find
\begin{equation}
\Delta R=\JMS^2 \pi ^2(-\tilde \tau +\tilde \tau ^2)v^2+O(v^3).
\end{equation}

\subsection{Geometry at finite $\lambda $} \label{sec:finite_lambda}

In this section we attempt to assign a meaning to the geometry for finite values of $\lambda $, and compare it to the semiclassical expansion we found.

Naively, given the 2-point function $G$, we can express $\langle e^g\rangle$ as $G^{\lambda /\tilde \lambda } $ in the limit of taking an operator made of a single fermion. Since the latter expression depends on the operator size through $\tilde q$, that would suggest to take
\begin{equation}
\langle e^g\rangle =\lim _{\tilde q \to 1} G^{\lambda /\tilde \lambda } .
\end{equation}

Using this, the curvature corresponding to the expectation value of the metric turns out to have a rather simple physical interpretation as the fluctuations in the energy of the operators in the probe limit (the limit of light operators). Let us introduce the following notation. Consider the unnormalized 2-point function in the energy representation, where we integrate over the energies $E_1$ and $E_2$ in the two intervals between the two operators. We denote by $\langle (E_1-E_2) \rangle _{\text{2-pt func}} $ the same representation with the insertion of $(E_1-E_2)$ in the integral. With this notation, $\langle 1\rangle _{\text{2-pt func}} $ is simply the unnormalized 2-point function. When considering the curvature, we should take a product of such expressions and we will take the limit $\tilde q \to 1$ after writing the product. In this case, the curvature is given by
\begin{equation}
R = - \lim _{\tilde q \to 1} G^{-\log q/\log \tilde q } \cdot  \frac{\log q}{\log \tilde q} \left[ \frac{\langle (E_1-E_2)^2 \rangle _{\text{2-pt func}} }{\langle 1 \rangle _{\text{2-pt func}} } -\left( \frac{\langle E_1-E_2\rangle _{\text{2-pt func}} }{\langle 1\rangle_{\text{2-pt func}}  } \right) ^2\right] .
\end{equation}
This expression indeed represents fluctuations in the energy.

\begin{figure}[h]
\centering
\includegraphics[width=1\textwidth]{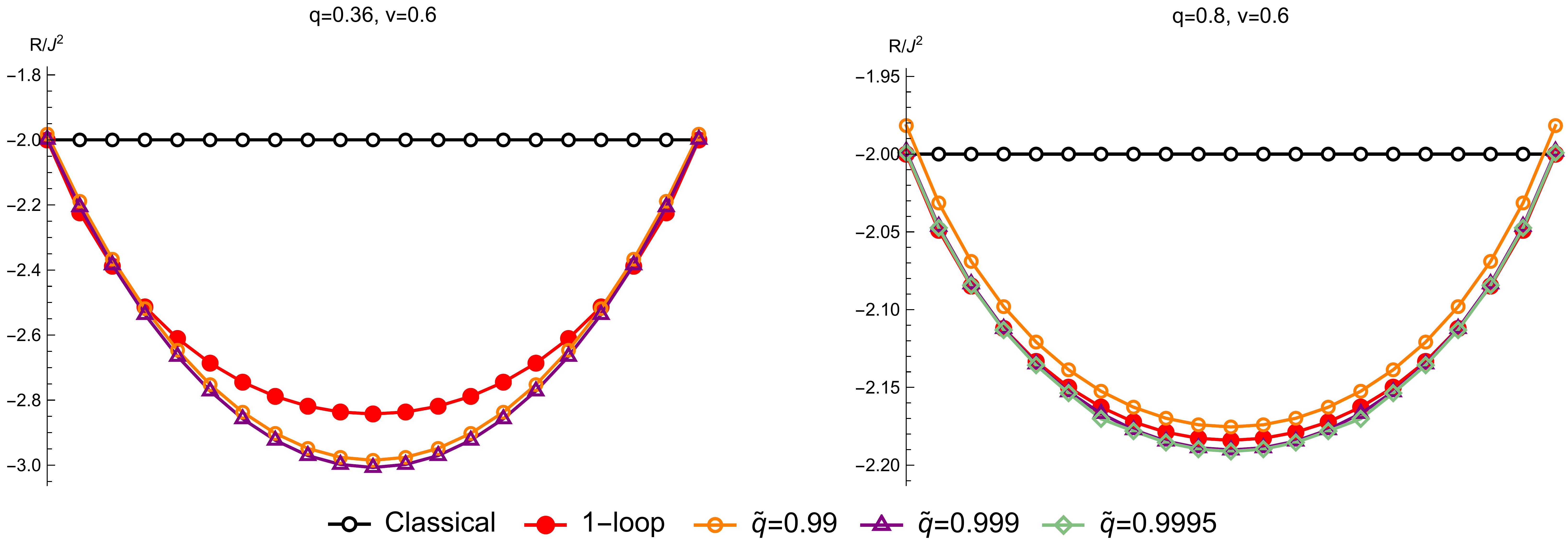}
\caption{Plot of the curvature $R/\JMS^2$ as a function of $\tau \JMS$ for $v=0.6$, and two values of $q$: $q=0.36$ (left) and $q=0.8$ (right).}
\label{fig:Ricci_v_06}
\end{figure}

A plot of the curvature for $v=0.6$ is shown in Fig.\ \ref{fig:Ricci_v_06}. The classical curvature is constant. The 1-loop result based on the formula \eqref{eq:Ricci_1_loop} is also shown in the figure. Lastly, we show a numerical evaluation of the curvature for several values of $\tilde q$, and we remember that we should take $\tilde q \to 1$. It seems that we get a nice convergence to the value of $\tilde q \to 1$. We can now compare our 1-loop result to the numerical value. We see that for $q=0.36$, corresponding to the expansion parameter $\lambda \approx 1$ the two results are not far. The other value of $q$ we show is $q=0.8$, which corresponds to $\lambda \approx 0.2$. Here indeed the approximation is very good, providing another test of our 1-loop computation.

\begin{figure}[h]
\centering
\includegraphics[width=1\textwidth]{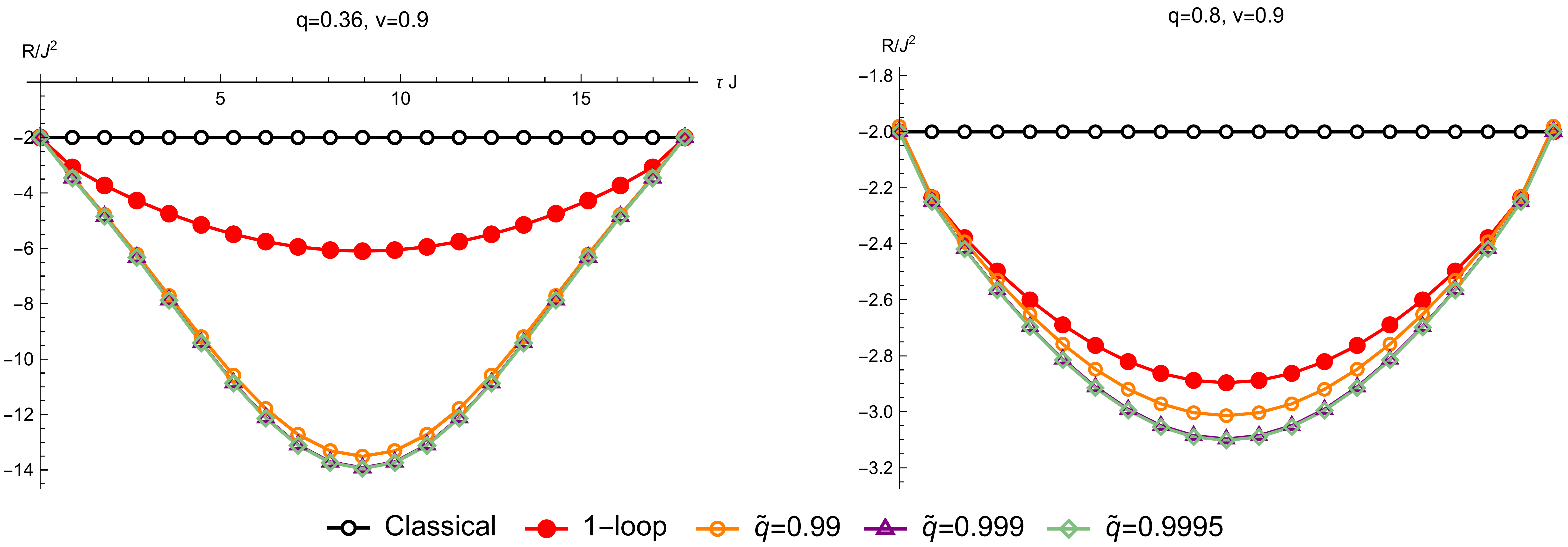}
\caption{Plot of the curvature $R/\JMS^2$ as a function of $\tau \JMS$ for $v=0.9$, and two values of $q$: $q=0.36$ (left) and $q=0.8$ (right).}
\label{fig:Ricci_v_09}
\end{figure}

We can also understand now how good the approximation is not only as a function of $q$ but also as a function of the temperature. A lower value of the temperature, corresponding to $v=0.9$ is plotted in Fig.\ \ref{fig:Ricci_v_09}. As expected, the 1-loop approximation becomes better as $q \to 1$, but we learn that the approximation becomes less and less accurate as we go to lower temperatures. Indeed, we see that for $q=0.36$ and $v=0.9$, the 1-loop computation is quite far from the actual result.

\section{Partially entangled thermal states} \label{sec:PETS}

A class of partially entangled states, generalizing the thermofield double, can be constructed using a local operator $\cO $. They are defined on two copies of a Hilbert space, and are given by
\begin{equation}
|\Psi\rangle  = \sum _{n,m} e^{-\frac{\beta _R}{2} E_m} e^{-\frac{\beta _L}{2} E_n} \cO _{mn} |\bar n\rangle _L |m\rangle _R
\end{equation}
where $\cO _{mn} =\langle m|\cO |n\rangle $ and $|n\rangle $ is an energy basis. $\beta _L$ and $\beta _R$ are parameters of the state. Instead of looking at states defined on two copies of the Hilbert space, we can consider equivalently operators on a single copy of the Hilbert space, and then these states are represented by
$ e^{-\frac{\beta _R}{2} H} \cO e^{-\frac{\beta _L}{2} H} $ so that
\begin{equation}
{}_L \langle \psi_1| {}_R \langle \psi_2| \Psi\rangle =\langle \psi_2 |e^{-\frac{\beta _R}{2} H} \cO e^{-\frac{\beta _L}{2} H} |\psi_1^*\rangle .
\end{equation}
We can thus obtain these states by doing a Euclidean path integral over a segment of length $\frac{\beta _L}{2} $, inserting the operator $\cO $, and finally having another interval of size $\frac{\beta _R}{2} $:
\begin{equation}
|\Psi\rangle  \cong \quad
    \tikz[baseline=-0.5ex]{
    \draw[thick] (1,0) arc (180:300:1) node (a)[pos=0]{} node(b)[pos=1]{} node[pos=0.5,above]{{\footnotesize $\frac{\beta _L}{2} $}};
    \draw[thick] (b) arc (230:360:1.12) node (c)[pos=1]{} node[pos=0.5,above]{{\footnotesize $\frac{\beta _R}{2} $}};
    \node at (b) [circle,fill,inner sep=1.5pt,color=red]{};
    \node at (a) [circle,fill,inner sep=1pt]{};
    \node at (c) [circle,fill,inner sep=1pt]{};
    } \quad .
\end{equation}
For the choice $\cO =1$, this state is the thermofield double with $\beta =\beta _L+\beta _R$. These states are called partially entangled thermal states (PETS) \cite{Goel:2018ubv}.

\subsection{Entanglement entropy}

To study the entanglement entropy, we should consider the reduced density matrix to the (say) right subsystem. It is given by
\begin{equation}
\begin{split}
\rho  &= \tr _L |\Psi\rangle \langle \Psi| = \sum _{m,n,k} e^{-\frac{\beta _R}{2} E_m} \cO _{mn} e^{-\beta _LE_n} \cO ^{\dagger} _{nk} e^{-\frac{\beta _R}{2} E_k} |m\rangle \langle k|=\\
& = e^{-\frac{\beta _R}{2} H} \cO e^{-\beta _LH} \cO ^{\dagger} e^{-\frac{\beta _R}{2} H} . 
\end{split}
\end{equation}
This density matrix is given by the Euclidean path integral represented by the following diagram
\begin{equation}
\rho  = \quad
    \tikz[baseline=-0.5ex]{
    \draw[thick] (1,0) arc (180:300:1) node (a)[pos=0]{} node(b)[pos=1]{} node[pos=0,right]{{\footnotesize $\beta _L $}};
    \draw[thick] (b) arc (230:355:1.12) node (c)[pos=1]{} node[pos=0.5,above]{{\footnotesize $\frac{\beta _R}{2} $}};
    \node at (b) [circle,fill,inner sep=1.5pt,color=red]{};
    \draw[thick] (1,0) arc (180:60:1) node(bb)[pos=1]{};
     \draw[thick] (bb) arc (130:5:1.12) node (cc)[pos=1]{} node[pos=0.5,below]{{\footnotesize $\frac{\beta _R}{2} $}};
    \node at (bb) [circle,fill,inner sep=1.5pt,color=red]{};
    \node at (b)[above]{{\footnotesize $\cO $}};
    \node at (bb)[below]{{\footnotesize $\cO ^{\dagger}  $}};
    \node at (c) [circle,fill,inner sep=1pt]{};
    \node at (cc) [circle,fill,inner sep=1pt]{};
    } \quad .
\end{equation}
Note that in our case the operators we consider are Hermitian.

We are interested in computing the Renyi entropies for PETS. A similar quantity is the modular entropy that we will denote by $S_n$ for simplicity, defined by
\begin{equation} \label{eq:mod_Renyi_entropy}
S_n= - n^2 \pder{}{n} \left[ \frac{1}{n} \log Z_n\right] =\left( 1-n \pder{}{n} \right) \log Z_n
,\qquad Z_n=\tr \rho ^n .
\end{equation}
It is equivalent to computing the Renyi entropy, and has a more direct holographic interpretation \cite{Lewkowycz:2013nqa,Dong:2016fnf}. Note that we can use an unnormalized density matrix in this equation, and it does not affect the result for $S_n$ (since it adds an $n$-independent constant inside the derivative).
For the same reason, $S_n$ is not affected by the normalization of the operator.
We will still refer to $S_n$ as the Renyi entropy for convenience. The entanglement entropy is still given by $S=\lim _{n \to 1} S_n$.

We consider PETS with the parametrization
\begin{equation}
\begin{split}
& \tau  = \frac{\beta _R}{2} ,\\
& \beta =\beta _L+\beta _R.
\end{split}
\end{equation}
With this parametrization, $Z_1$ is the 2-point function in a thermal circle of size $\beta $ where the distance between the operators is $2\tau $ (rather than $\tau $; we will use this convention following \cite{Goel:2018ubv}). The $n$'th Renyi entropy is then given by a $2n$-point function where the thermal circle is of size $n\beta $. Note, importantly, that we should use unnormalized correlation functions (rather than correlators normalized by the partition function).

\subsection{Light operators}

For the size of the operator, we use the notation mentioned previously
\begin{equation}
l=\frac{\tilde \lambda }{\lambda } .
\end{equation}
In this section we consider light operators, that is, small $\tilde \lambda $. In the $2n$-point correlation functions, there are different contractions of operators that could occur, since we have $2n$ insertions of the same operator. (For more details on operators in double-scaled SYK, see \cite{Berkooz:2018jqr}.) The simplest contraction is to pair each two consecutive operators of distance $2\tau $ part. For $n=4$ this is shown on the LHS of Fig.\ \ref{fig:n_4_two_contractions}. This configuration is studied in appendix \ref{sec:2n_point_function}. For light operators, we have large $N$ factorization, and the correlation function is given by a product of $n$ terms. Each contraction is proportional to
\begin{equation} \label{eq:contraction1}
\frac{1}{\cos \left( \pi v_n\left( \frac{1}{2} -\frac{2\tau }{n\beta }\right) \right)  ^{2l}} 
\end{equation}
where $v_n$ is the $v$ corresponding to $n \beta$, rather than $\beta$.
Similarly, we can contract each operator with the operator on the other side of it, and then each contraction is assigned
\begin{equation} \label{eq:contraction2}
\frac{1}{\cos \left( \pi v_n\left( \frac{1}{2} -\frac{\beta -2\tau }{n\beta }\right) \right)  ^{2l}} .
\end{equation}
This is shown on the RHS of Fig.\ \ref{fig:n_4_two_contractions}.
\begin{figure}[h]
\centering
\includegraphics[width=0.6\textwidth]{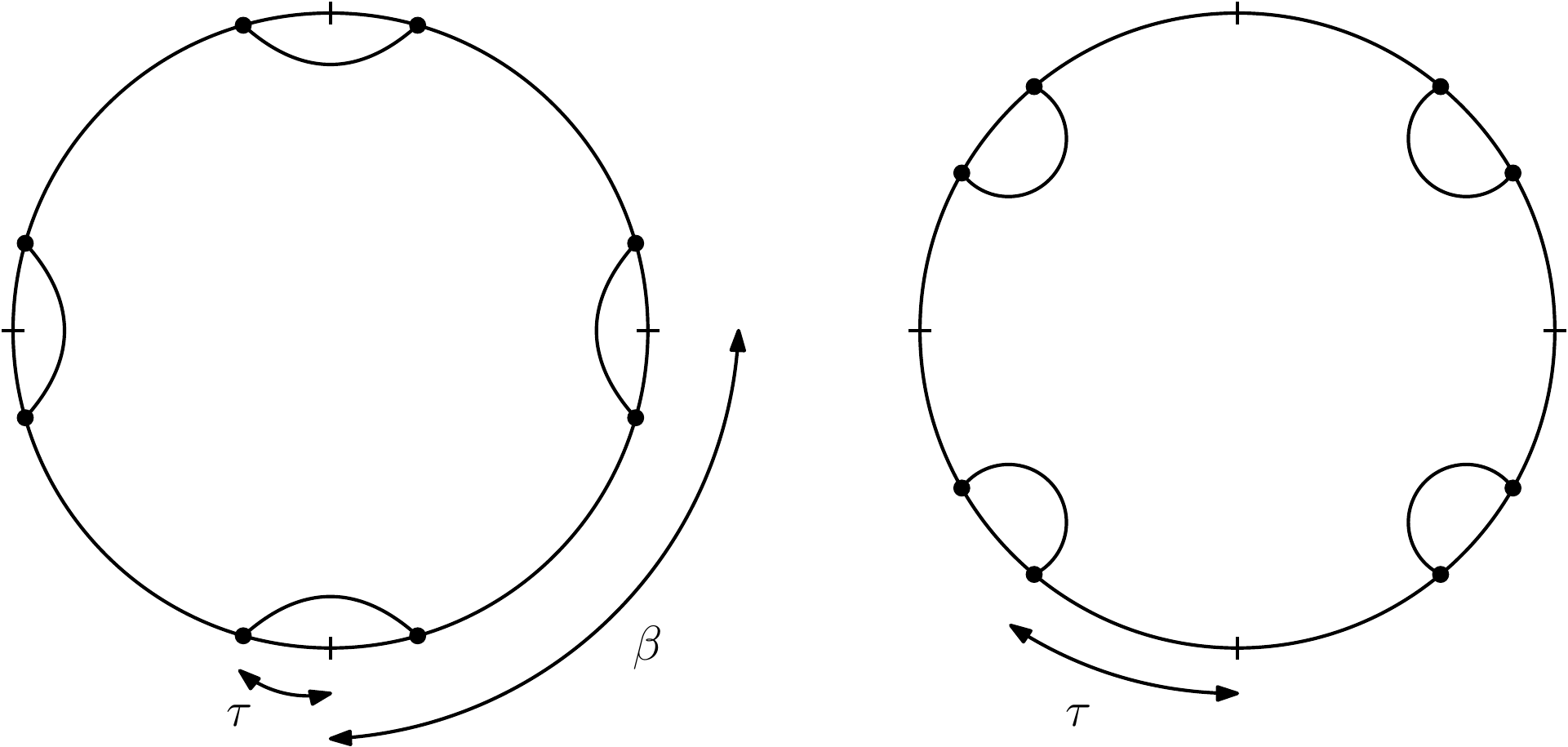}
\caption{Two possible contractions in the case $n=4$.}
\label{fig:n_4_two_contractions}
\end{figure}

For $0<\tau <\frac{\beta }{4} $, \eqref{eq:contraction1} is more dominant than \eqref{eq:contraction2}, and for $\frac{\beta }{4} <\tau <\frac{\beta }{2} $ (the maximal value of $2\tau $ is $\beta $) the latter is more dominant. This is similar to \cite{Goel:2018ubv}, and importantly it is unaffected by the fact that we have the $v$ dependence here. This behavior is intuitive, as we simply contract the closest operators. Considering large $l$, the other contractions are exponentially suppressed in $l$.

In the first case, $0<\tau <\frac{\beta }{4} $, the unnormalized correlation function is given by the following, where now we restore the $N$ dependence that was absent before due to the normalization of the trace,
\begin{equation}
Z_n = \exp\Bigg\{\frac{N}{2} \log 2 -\frac{\pi ^2v_n^2}{2\lambda } +\frac{2\pi v_n}{\lambda } \tan \frac{\pi v_n}{2} + 2nl \left[ \log\cos \frac{\pi v_n}{2} -\log \cos \left( \pi v_n\left( \frac{1}{2} -\frac{2\tau }{n\beta } \right) \right) \right] \Bigg\} .
\end{equation}
For $\frac{\beta }{4} <\tau <\frac{\beta }{2} $,
\begin{equation}
Z_n = \exp\Bigg\{\frac{N}{2} \log 2 -\frac{\pi ^2v_n^2}{2\lambda } +\frac{2\pi v_n}{\lambda } \tan \frac{\pi v_n}{2} + 2nl \left[ \log\cos \frac{\pi v_n}{2} -\log \cos \left( \pi v_n\left( \frac{1}{2} -\frac{\beta-2\tau }{n\beta } \right) \right) \right] \Bigg\} .
\end{equation}
Using \eqref{eq:mod_Renyi_entropy}, we find for the Renyi entropies
\begin{equation}
\begin{split}
& S_n = \frac{N}{2} \log 2+\frac{\pi v_n}{1+\frac{\pi v_n}{2} \tan \frac{\pi v_n}{2} } \Bigg[ -\frac{\pi v_n}{2\lambda } +\left( -\frac{\pi ^2v_n^2}{4\lambda } +ln\right) \tan \frac{\pi v_n}{2} -\\
& \qquad -l\left( n+\pi v_n \frac{\tau _{\text{min}}}{\beta } \tan \frac{\pi v_n}{2} \right) \tan\left( \pi v_n\left( \frac{1}{2} -\frac{\tau _{\text{min}}}{n\beta } \right) \right) \Bigg],
\end{split}
\end{equation}
where
\begin{equation}
\tau _{\text{min}}  = \min\left( 2\tau ,\beta -2\tau \right) .
\end{equation}

If we restrict to small temperatures, keeping the first subleading order in $1/(\beta \JMS)$, we find that $v_n \approx 1-\frac{2}{n\beta \JMS} $. Keeping $\tau /\beta $ fixed, we find in this limit
\begin{equation} \label{eq:Renyi_ent_light_ops_low_temp}
S_n = \frac{N}{2} \log 2-\frac{\pi ^2}{2\lambda }+\frac{2\pi ^2}{n\lambda \beta \JMS}  +2l\left( n-\frac{\pi \tau _{\text{min}} }{\beta } \cot \frac{\pi \tau _{\text{min}} }{n\beta} \right) .
\end{equation}
Clearly the expressions should be symmetric under exchanging the left and right Hilbert spaces, which corresponds to taking $2\tau  \to \beta -2\tau $, and indeed this is the case.
Recalling the relation $C=1/(2\lambda \JMS)$ \cite{Berkooz:2018jqr} to the Schwarzian coupling $C$, \eqref{eq:Renyi_ent_light_ops_low_temp} reproduces the result in \cite{Goel:2018ubv}.

\subsection{Heavy operators}

Now we do not assume that $\tilde \lambda $ is small. Just as before, we have two cases. For $0<\tau <\frac{\beta }{4} $, the dominant contraction is the one on the LHS of Fig.\ \ref{fig:n_4_two_contractions}. This configurations has a $\mathbb{Z} _n$ replica symmetry. This setup is studied in appendix \ref{sec:2n_point_function}. Assuming there is no replica symmetry breaking, at the saddle point the solution has $\theta _1=\theta _2=\cdots =\theta _n$ in the notation of appendix \ref{sec:2n_point_function}. The action then takes the form
\begin{equation}
\log Z_n =\frac{N}{2} \log 2 -\frac{\pi ^2}{6\lambda } + \li_2\left( e^{\pm 2i\phi } \right) +n \tilde I(\tau ,\beta ,\theta ,\phi )
\end{equation}
for some function $\tilde I$.
In calculating $n \pder{}{n} \log Z_n$, the saddle point values of $\theta $ and $\phi $ depend on $n$, but at the saddle point solution the derivative with respect to these variables vanishes. We should therefore consider only the explicit derivative with respect to $n$, leading to
\begin{equation}
S_n = \frac{N}{2} \log 2-\frac{\pi ^2}{2\lambda } +\frac{2\phi }{\lambda } (\pi -\phi ).
\end{equation}
The value of $\phi $ (that depends on $n$, $l$, $\tau $, and $\beta $) is determined by the saddle point equations of appendix \ref{sec:2n_point_function}, which in the current conventions are
\begin{equation} \label{eq:heavy_ops_saddle_point}
\begin{split}
& 2\theta -\pi +\arctan\left[ \frac{\sin(\theta +\phi  )}{e^{\tilde \lambda } -\cos(\theta +\phi  )} \right] +\arctan\left[ \frac{\sin(\theta -\phi  )}{e^{\tilde \lambda } -\cos(\theta -\phi  )} \right] =2\JMS \tau \sin\theta , \\[3mm]
& \frac{2\phi  -\pi}{n} + \arctan\left[ \frac{\sin(\theta +\phi  )}{e^{\tilde \lambda } -\cos(\theta +\phi  )} \right] -\arctan\left[ \frac{\sin(\theta -\phi  )}{e^{\tilde \lambda } -\cos(\theta -\phi  )} \right]  =\JMS\left( \beta -2\tau  \right) \sin \phi  .
\end{split}
\end{equation}
The case $\frac{\beta }{4} <\tau <\frac{\beta }{2} $ is the same, replacing $2\tau  \to \beta -2\tau $. For heavy operators, we can only provide the solution in this implicit form. At low energies, following \cite{Berkooz:2018jqr}, we can define $\phi =\pi -\lambda \tilde p$ and $\theta =\pi -\lambda \tilde k$ and see that at small $\lambda $ \eqref{eq:heavy_ops_saddle_point} reproduces the result in \cite{Goel:2018ubv}.

\section{Discussion}

In this paper we studied a semiclassical expansion in double-scaled SYK, controlled by a parameter $\lambda $. The observables of the large $p$ SYK model (at least the ones we considered) are given essentially trivially by the saddle point values of the results of double-scaled SYK. The 1-loop correction is more intricate, and we have studied it as well, and compared the analytic expression we found to numerical evaluation. We considered the two dimensional geometry that corresponds to these results. We have seen that classically the geometry can be interpreted using AdS or dS.\footnote{A relation of double-scaled SYK to de Sitter space was suggested in \cite{HermanDeSitter,Susskind:2021esx,Susskind:2022dfz,Lin:2022nss,Susskind:2022bia,Rahman:2022jsf}.}

There are several interesting future directions.
The first direction is to extend these results to other models with global symmetry and supersymmetry \cite{Berkooz:2020xne,Berkooz:2020uly}. Whenever there are global symmetries, we have additional gauge fields in the bulk, and it would be interesting to understand them in the formalism that we have here.

As mentioned before, one can expect to be able to discuss usual semiclassical geometry when doing an asymptotic expansion in $\lambda $. The situation for finite $\lambda $ is less clear, and we only commented on it in section \ref{sec:finite_lambda}. Recently a non-commutative dual of double-scaled SYK has been described in \cite{Berkooz:2022mfk}. It would be desirable to understand if there is any relation between such a gravitational description and the formalism used here.

The quantities studied here are averaged single trace observables and correspond to a single boundary. We have not described what happens when there are several boundaries \cite{Pluma:2019pnc,Jia:2019orl,Berkooz:2020fvm}. In JT gravity, there are wormhole contributions in such cases.

Finally, while we do not know of a direct relation between our description and JT gravity, it is desirable to understand if such a connection exists. Moreover, we find corrections to the AdS geometry, and it is natural to look for a deformation of JT gravity that leads to this geometry. An additional question is whether a kinematic space construction can lead to the geometries we found. The volume form of kinematic space is related to geodesic lengths in the simplest description. However, one can replace geodesics by extremal curves of an action that generalizes the simplest action corresponding to the length of a curve. In that case, a different volume form and metric can be found, and it is interesting to understand whether there is such a description that leads to the geometry we found, at least at 1-loop.

\section*{Acknowledgments}

We thank Ahmed Almheiri, Micha Berkooz, and Juan Maldacena for useful discussions. We also thank Silviu Pufu and Bernardo Zan for useful help with the numerics. The research of VN and HV is supported by NSF grant PHY-2209997.

\appendix

\section{Uncrossed $2n$-point functions} \label{sec:2n_point_function}

\begin{figure}[h]
\centering
\includegraphics[width=0.4\textwidth]{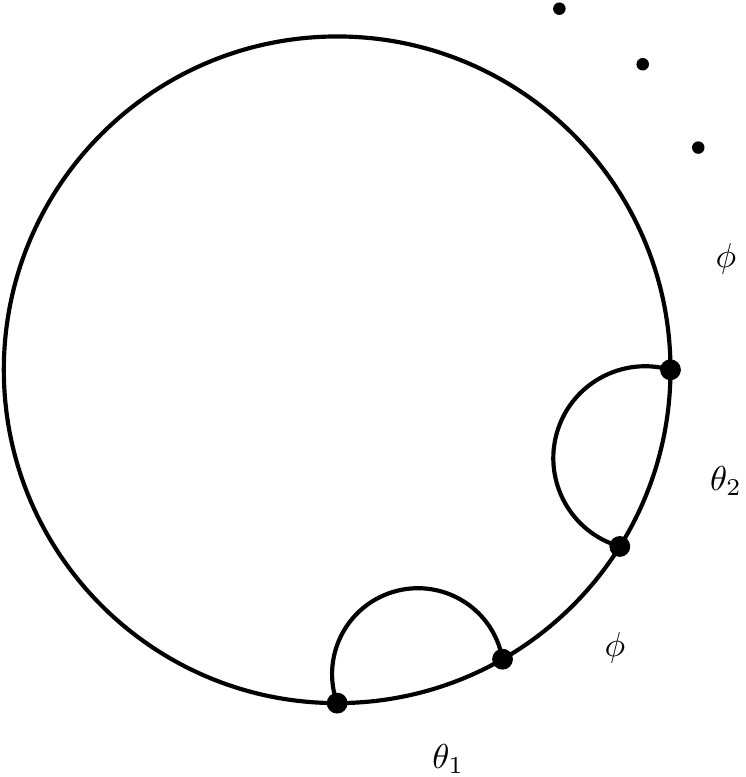}
\caption{Uncrossed $2n$-point function.}
\label{fig:2n_point_function}
\end{figure}

In this appendix we consider the extension of the calculation of the 2-point function to a $2n$-point function, where the operators come consecutively in pairs, that is the operator flavors are such that we arrange $ \cO _1 \cO _1 \cO _2 \cO _2 \cdots \cO _n \cO _n$. The times are such that the time difference between each pair is $\tau _j$ where $j=1,\cdots ,n$ and the total Euclidean time is $\beta $. See Fig.\ \ref{fig:2n_point_function}. Here we consider only the leading order of this correlation function.

In this setup, we have one integration variable $\theta _j$ for every pair of operators, and there is a single variable $\phi $ that runs between the pairs (it is the same variable in all intermediate regions). There is a parameter $\tilde q_j$ for every pair, determined by the flavor of the operators, and again we define $\tilde q_j = e^{-\tilde \lambda _j} $. The form of the correlation function for small $\lambda $ is
\begin{equation}
(q;q)_{\infty } ^{n+1} (\tilde q^2;q)_{\infty } ^n \int d\theta _1 \cdots d\theta _n d \phi \,   e^{-\frac 1 \lambda {f(\theta_1,..,\theta_n,\phi)}  } 
\end{equation}
where now
\begin{equation}
\begin{split}
& f= \li_2\left( e^{\pm 2i\phi  } \right) + \sum _{j=1} ^n \left(\li _2\left( e^{\pm 2i\theta _j} \right) - \li_2\bigl( e^{-\tilde \lambda _j+i(\pm \theta _j \pm \phi  )} \bigr) +2 \JMS \tau _j\cos \theta _j\right)+2\JMS \Bigl(\beta -\sum _j\tau _j\Bigr) \cos \phi  .
\end{split}
\end{equation}

The saddle point equations are as follows. For $j=1,\cdots ,n$ we have
\begin{equation}
2\theta _j-\pi +\arctan\left[ \frac{\sin(\theta _j+\phi  )}{e^{\tilde \lambda_j } -\cos(\theta _j+\phi  )} \right] +\arctan\left[ \frac{\sin(\theta _j-\phi  )}{e^{\tilde \lambda_j } -\cos(\theta _j-\phi  )} \right] =\JMS \tau _j\sin\theta _j
\end{equation}
and the saddle point equation for $\phi  $ is
\begin{equation}
2\phi  -\pi +\sum _{j=1} ^n \left( \arctan\left[ \frac{\sin(\theta _j+\phi  )}{e^{\tilde \lambda_j } -\cos(\theta _j+\phi  )} \right] -\arctan\left[ \frac{\sin(\theta _j-\phi  )}{e^{\tilde \lambda_j } -\cos(\theta _j-\phi  )} \right] \right) =\JMS\biggl( \beta -\sum _j \tau _j \biggr) \sin \phi  .
\end{equation}

Considering now small $\tilde \lambda _j$, we take a similar ansatz to before
\begin{equation}
\begin{split}
& \phi  =\theta _0,\\
& \theta _j = \theta _0+\alpha _j\tilde \lambda _j .
\end{split}
\end{equation}
The saddle point equations now become
\begin{equation}
\begin{split}
&\theta _0-\frac{\pi }{2} +\arctan \alpha _j = \JMS \tau _j \sin \theta _0 ,\\
& (2-n)\left( \theta _0-\frac{\pi }{2} \right) -\sum _j \arctan \alpha _j = \JMS \biggl( \beta -\sum _j \tau _j\biggr) \sin \theta _0.
\end{split}
\end{equation}
Just as before, if we define $\theta _0 = \frac{\pi }{2} +\frac{\pi v}{2} $, we get the same relation \eqref{eq:beta_v_rel}. The remaining equations fix
\begin{equation}
\alpha _j = -\tan\left[ \frac{\pi v}{2} \Bigl( 1-\frac{2\tau _j}{\beta } \Bigr) \right] .
\end{equation}
Plugging back into the correlator, we find up to the order we work in
\begin{equation}
\begin{split}
&f = -\left( \frac{1}{6} +\frac{n}{3} \right) \pi ^2+\frac{\pi ^2v^2}{2} -2\pi v\tan \frac{\pi v}{2} +\\[1mm] &\qquad +\sum _j \tilde \lambda _j \Bigg[ 2-2\log(2\tilde \lambda _j)- 2\log \cos \frac{\pi v}{2} +2\log \cos \left( \pi v\Bigl( \frac{1}{2} -\frac{\tau _j}{2} \Bigr) \right) \Bigg].
\end{split}
\end{equation}
This results in the correlation function
\begin{equation}
G = \prod _{j=1} ^n \left[ \frac{\cos \frac{\pi v}{2} }{\cos \left( \pi v\Bigl( \frac{1}{2} -\frac{\tau _j}{\beta } \Bigr) \right) } \right] ^{2\tilde \lambda _j/\lambda } .
\end{equation}
This result is just the product of 2-point functions, which is the leading order behavior.

\bibliography{main}

\providecommand{\href}[2]{#2}\begingroup\raggedright\begin{thebibliography}{10}

\bibitem{Erdos:2014zgc}
L.~Erd\H{o}s and D.~Schr\"oder, \emph{{Phase Transition in the Density of
  States of Quantum Spin Glasses}},
  \href{https://doi.org/10.1007/s11040-014-9164-3}{\emph{Math. Phys. Anal.
  Geom.} {\bfseries 17} (2014) 441}
  [\href{https://arxiv.org/abs/1407.1552}{{\ttfamily 1407.1552}}].

\bibitem{Cotler:2016fpe}
J.S.~Cotler, G.~Gur-Ari, M.~Hanada, J.~Polchinski, P.~Saad, S.H.~Shenker
  et~al., \emph{{Black Holes and Random Matrices}},
  \href{https://doi.org/10.1007/JHEP05(2017)118}{\emph{JHEP} {\bfseries 05}
  (2017) 118} [\href{https://arxiv.org/abs/1611.04650}{{\ttfamily
  1611.04650}}].

\bibitem{Berkooz:2018qkz}
M.~Berkooz, P.~Narayan and J.~Simon, \emph{{Chord diagrams, exact correlators
  in spin glasses and black hole bulk reconstruction}},
  \href{https://doi.org/10.1007/JHEP08(2018)192}{\emph{JHEP} {\bfseries 08}
  (2018) 192} [\href{https://arxiv.org/abs/1806.04380}{{\ttfamily
  1806.04380}}].

\bibitem{Berkooz:2018jqr}
M.~Berkooz, M.~Isachenkov, V.~Narovlansky and G.~Torrents, \emph{{Towards a
  full solution of the large N double-scaled SYK model}},
  \href{https://doi.org/10.1007/JHEP03(2019)079}{\emph{JHEP} {\bfseries 03}
  (2019) 079} [\href{https://arxiv.org/abs/1811.02584}{{\ttfamily
  1811.02584}}].

\bibitem{Maldacena:2016hyu}
J.~Maldacena and D.~Stanford, \emph{{Remarks on the Sachdev-Ye-Kitaev model}},
  \href{https://doi.org/10.1103/PhysRevD.94.106002}{\emph{Phys. Rev. D}
  {\bfseries 94} (2016) 106002}
  [\href{https://arxiv.org/abs/1604.07818}{{\ttfamily 1604.07818}}].

\bibitem{Sachdev:1992fk}
S.~Sachdev and J.~Ye, \emph{{Gapless spin fluid ground state in a random,
  quantum Heisenberg magnet}},
  \href{https://doi.org/10.1103/PhysRevLett.70.3339}{\emph{Phys. Rev. Lett.}
  {\bfseries 70} (1993) 3339}
  [\href{https://arxiv.org/abs/cond-mat/9212030}{{\ttfamily
  cond-mat/9212030}}].

\bibitem{Kitaev_talk}
A.~Kitaev, \emph{{A simple model of quantum holography}},
  \url{http://online.kitp.ucsb.edu/online/entangled15/kitaev/},
  \url{http://online.kitp.ucsb.edu/online/entangled15/kitaev2/}.

\bibitem{Polchinski:2016xgd}
J.~Polchinski and V.~Rosenhaus, \emph{{The Spectrum in the Sachdev-Ye-Kitaev
  Model}}, \href{https://doi.org/10.1007/JHEP04(2016)001}{\emph{JHEP}
  {\bfseries 04} (2016) 001}
  [\href{https://arxiv.org/abs/1601.06768}{{\ttfamily 1601.06768}}].

\bibitem{Mertens:2017mtv}
T.G.~Mertens, G.J.~Turiaci and H.L.~Verlinde, \emph{{Solving the Schwarzian via
  the Conformal Bootstrap}},
  \href{https://doi.org/10.1007/JHEP08(2017)136}{\emph{JHEP} {\bfseries 08}
  (2017) 136} [\href{https://arxiv.org/abs/1705.08408}{{\ttfamily
  1705.08408}}].

\bibitem{Mertens:2022irh}
T.G.~Mertens and G.J.~Turiaci, \emph{{Solvable Models of Quantum Black Holes: A
  Review on Jackiw-Teitelboim Gravity}},
  \href{https://arxiv.org/abs/2210.10846}{{\ttfamily 2210.10846}}.

\bibitem{Streicher:2019wek}
A.~Streicher, \emph{{SYK Correlators for All Energies}},
  \href{https://doi.org/10.1007/JHEP02(2020)048}{\emph{JHEP} {\bfseries 02}
  (2020) 048} [\href{https://arxiv.org/abs/1911.10171}{{\ttfamily
  1911.10171}}].

\bibitem{Choi:2019bmd}
C.~Choi, M.~Mezei and G.~S\'arosi, \emph{{Exact four point function for large
  $q$ SYK from Regge theory}},
  \href{https://doi.org/10.1007/JHEP05(2021)166}{\emph{JHEP} {\bfseries 05}
  (2021) 166} [\href{https://arxiv.org/abs/1912.00004}{{\ttfamily
  1912.00004}}].

\bibitem{Berkooz:2022mfk}
M.~Berkooz, M.~Isachenkov, P.~Narayan and V.~Narovlansky, \emph{{Quantum
  groups, non-commutative $AdS_2$, and chords in the double-scaled SYK model}},
   \href{https://arxiv.org/abs/2212.13668}{{\ttfamily 2212.13668}}.

\bibitem{Czech:2014ppa}
B.~Czech and L.~Lamprou, \emph{{Holographic definition of points and
  distances}}, \href{https://doi.org/10.1103/PhysRevD.90.106005}{\emph{Phys.
  Rev. D} {\bfseries 90} (2014) 106005}
  [\href{https://arxiv.org/abs/1409.4473}{{\ttfamily 1409.4473}}].

\bibitem{Czech:2015qta}
B.~Czech, L.~Lamprou, S.~McCandlish and J.~Sully, \emph{{Integral Geometry and
  Holography}}, \href{https://doi.org/10.1007/JHEP10(2015)175}{\emph{JHEP}
  {\bfseries 10} (2015) 175}
  [\href{https://arxiv.org/abs/1505.05515}{{\ttfamily 1505.05515}}].

\bibitem{Czech:2016xec}
B.~Czech, L.~Lamprou, S.~McCandlish, B.~Mosk and J.~Sully, \emph{{A
  Stereoscopic Look into the Bulk}},
  \href{https://doi.org/10.1007/JHEP07(2016)129}{\emph{JHEP} {\bfseries 07}
  (2016) 129} [\href{https://arxiv.org/abs/1604.03110}{{\ttfamily
  1604.03110}}].

\bibitem{Lam:2018pvp}
H.T.~Lam, T.G.~Mertens, G.J.~Turiaci and H.~Verlinde, \emph{{Shockwave S-matrix
  from Schwarzian Quantum Mechanics}},
  \href{https://doi.org/10.1007/JHEP11(2018)182}{\emph{JHEP} {\bfseries 11}
  (2018) 182} [\href{https://arxiv.org/abs/1804.09834}{{\ttfamily
  1804.09834}}].

\bibitem{Maldacena:2016upp}
J.~Maldacena, D.~Stanford and Z.~Yang, \emph{{Conformal symmetry and its
  breaking in two dimensional Nearly Anti-de-Sitter space}},
  \href{https://doi.org/10.1093/ptep/ptw124}{\emph{PTEP} {\bfseries 2016}
  (2016) 12C104} [\href{https://arxiv.org/abs/1606.01857}{{\ttfamily
  1606.01857}}].

\bibitem{Kitaev:2018wpr}
A.~Kitaev and S.J.~Suh, \emph{{Statistical mechanics of a two-dimensional black
  hole}}, \href{https://doi.org/10.1007/JHEP05(2019)198}{\emph{JHEP} {\bfseries
  05} (2019) 198} [\href{https://arxiv.org/abs/1808.07032}{{\ttfamily
  1808.07032}}].

\bibitem{Goel:2018ubv}
A.~Goel, H.T.~Lam, G.J.~Turiaci and H.~Verlinde, \emph{{Expanding the Black
  Hole Interior: Partially Entangled Thermal States in SYK}},
  \href{https://doi.org/10.1007/JHEP02(2019)156}{\emph{JHEP} {\bfseries 02}
  (2019) 156} [\href{https://arxiv.org/abs/1807.03916}{{\ttfamily
  1807.03916}}].

\bibitem{Lewkowycz:2013nqa}
A.~Lewkowycz and J.~Maldacena, \emph{{Generalized gravitational entropy}},
  \href{https://doi.org/10.1007/JHEP08(2013)090}{\emph{JHEP} {\bfseries 08}
  (2013) 090} [\href{https://arxiv.org/abs/1304.4926}{{\ttfamily 1304.4926}}].

\bibitem{Dong:2016fnf}
X.~Dong, \emph{{The Gravity Dual of Renyi Entropy}},
  \href{https://doi.org/10.1038/ncomms12472}{\emph{Nature Commun.} {\bfseries
  7} (2016) 12472} [\href{https://arxiv.org/abs/1601.06788}{{\ttfamily
  1601.06788}}].

\bibitem{HermanDeSitter}
H.~Verlinde, \emph{Talks given at the QIQG5 conference at UCDavis, August 2029,
  Quantum Gravity in the Southern Cone VIII in Bariloche, Argentina, December
  2019, and the SRITP workshop Gauge Theories and Black Holes at Weizmann
  Institute, December 2019}.

\bibitem{Susskind:2021esx}
L.~Susskind, \emph{{Entanglement and Chaos in De Sitter Space Holography: An
  SYK Example}}, \href{https://doi.org/10.22128/jhap.2021.455.1005}{\emph{JHAP}
  {\bfseries 1} (2021) 1} [\href{https://arxiv.org/abs/2109.14104}{{\ttfamily
  2109.14104}}].

\bibitem{Susskind:2022dfz}
L.~Susskind, \emph{{Scrambling in Double-Scaled SYK and De Sitter Space}},
  \href{https://arxiv.org/abs/2205.00315}{{\ttfamily 2205.00315}}.

\bibitem{Lin:2022nss}
H.~Lin and L.~Susskind, \emph{{Infinite Temperature's Not So Hot}},
  \href{https://arxiv.org/abs/2206.01083}{{\ttfamily 2206.01083}}.

\bibitem{Susskind:2022bia}
L.~Susskind, \emph{{De Sitter Space, Double-Scaled SYK, and the Separation of
  Scales in the Semiclassical Limit}},
  \href{https://arxiv.org/abs/2209.09999}{{\ttfamily 2209.09999}}.

\bibitem{Rahman:2022jsf}
A.A.~Rahman, \emph{{dS JT Gravity and Double-Scaled SYK}},
  \href{https://arxiv.org/abs/2209.09997}{{\ttfamily 2209.09997}}.

\bibitem{Berkooz:2020xne}
M.~Berkooz, N.~Brukner, V.~Narovlansky and A.~Raz, \emph{{The double scaled
  limit of Super--Symmetric SYK models}},
  \href{https://doi.org/10.1007/JHEP12(2020)110}{\emph{JHEP} {\bfseries 12}
  (2020) 110} [\href{https://arxiv.org/abs/2003.04405}{{\ttfamily
  2003.04405}}].

\bibitem{Berkooz:2020uly}
M.~Berkooz, V.~Narovlansky and H.~Raj, \emph{{Complex Sachdev-Ye-Kitaev model
  in the double scaling limit}},
  \href{https://doi.org/10.1007/JHEP02(2021)113}{\emph{JHEP} {\bfseries 02}
  (2021) 113} [\href{https://arxiv.org/abs/2006.13983}{{\ttfamily
  2006.13983}}].

\bibitem{Pluma:2019pnc}
M.~Pluma and R.~Speicher, \emph{{A dynamical version of the SYK model and the
  q-Brownian motion}},
  \href{https://doi.org/10.1142/S2010326322500319}{\emph{Random Matrices:
  Theory and Applications} {\bfseries 11} (2022) 2250031}
  [\href{https://arxiv.org/abs/1905.12999}{{\ttfamily 1905.12999}}].

\bibitem{Jia:2019orl}
Y.~Jia and J.J.M.~Verbaarschot, \emph{{Spectral Fluctuations in the
  Sachdev-Ye-Kitaev Model}},
  \href{https://doi.org/10.1007/JHEP07(2020)193}{\emph{JHEP} {\bfseries 07}
  (2020) 193} [\href{https://arxiv.org/abs/1912.11923}{{\ttfamily
  1912.11923}}].

\bibitem{Berkooz:2020fvm}
M.~Berkooz, N.~Brukner, V.~Narovlansky and A.~Raz, \emph{{Multi-trace
  correlators in the SYK model and non-geometric wormholes}},
  \href{https://doi.org/10.1007/JHEP09(2021)196}{\emph{JHEP} {\bfseries 21}
  (2020) 196} [\href{https://arxiv.org/abs/2104.03336}{{\ttfamily
  2104.03336}}].

\end{thebibliography}\endgroup
\bibliographystyle{JHEP}
\end{document}